\documentstyle[epsfig]{article}
\input{tcilatex}

\begin{document}

\title{Algebraic Model for Quantum Scattering. \\
Reformulation, Analysis and Numerical Strategies}
\author{V. S. Vasilevsky \\
%EndAName
Bogolyubov Institute for Theoretical Physics, \\
Ukrainian Academy of Sciences, 252143, Kiev 143, Ukraine \\
\and F. Arickx \\
%EndAName
University of Antwerp \\
Universitair Centrum Antwerpen (RUCA)\\
Department of Mathematics and Computer Science\\
Groenenborgerlaan 171, B2020 Antwerpen, Belgium.}
\date{\today}
\maketitle

\begin{abstract}
The convergence problem for scattering states is studied in detail within
the framework of the Algebraic Model, a representation of the
Schr\"{o}dinger equation in an $L^2$ basis. The dynamical equations of this
model are reformulated featuring new ``Dynamical Coefficients'', which
explicitly reveal the potential effects. A general analysis of the Dynamical
Coefficients leads to an optimal basis yielding well converging, precise and
stable results. A set of strategies for solving the equations for
non-optimal bases is formulated based on the asymptotic behaviour of the
Dynamical Coefficients. These strategies are shown to provide a dramatically
improved convergence of the solutions.
\end{abstract}

%\listoffigures

%\bibliographystyle{unsrt}  

PACS number(s): 03.80, 03.65.N, 03.65.F

%\narrowtext

%\twocolumn

\section{Introduction}

In the quest for solving the Schr\"{o}dinger equation for both bound and
continuum states, square-integrable bases have been repeatedly used. For
bound states this turns out to be a traditional way to obtain the spectral
properties of quantum systems. It has been shown however that a single
representation of the Schr\"{o}dinger equation in terms of an $L^2$ basis
can be formulated that allows for a description of both bound and continuum
states \cite{kn:Heller1,kn:Yamani,kn:Fil_Okhr,kn:Fil,kn:Smirnov}.

A version of such a formulation is called the Algebraic Model of the
Resonating Group Method (hereafter referred to as the Algebraic Model or
AM). It was originally tailored to treat clusterized problems, but can be
used for all kinds of quantum mechanical many-particle configurations
without major modifications. The AM has been formulated in terms of
different types of bases, depending on the more specific features of the
quantum system considered. One very important feature of the AM is the fact
that the boundary conditions of the system are translated from a coordinate
space context to the context of expansion coefficients, and are explicitly
incorporated in the dynamical equations \cite
{kn:Heller1,kn:Yamani,kn:Fil_Okhr,kn:Fil}.

We will consider a specific AM formulation (i.e.\ a specific $L^2$ basis) to
elucidate the analysis of the method. The methodology used for this analysis
is of a general nature however, and can be repeated for other bases. The
specific AM version chosen in this paper features an oscillator basis, and
is in particular very suitable for obtaining both the bound and continuum
spectra of nuclear systems with very different configurational properties;
where appropriate, nuclear spectroscopic units will therefore be used. The
choice of this AM version is mainly due to the background of the authors.
Results of the AM approach considered here have already been reported on 
\cite{kn:fil_rev2,kn:alphacalc}, and show in particular how the coupling of
cluster and collective configurations can be treated seamlessly in such a
description.

Where necessary, a specific form for the potential operator will be used. We
will consider a Gaussian form in this work. Although this again is a popular
potential form in nuclear spectroscopic calculations, it is also an
interesting functional form to approximate a large variety of potentials by
discrete and continuous superpositions.

We will concentrate our attention to the solutions of the dynamical
equations for scattering situations only, as these are much more involved
than the bound state problems. Indeed, the latter can be well approximated
by a simple diagonalization of the energy matrix, as is well known.

We will specifically discuss how strongly the convergence of the solutions
of an AM system depend on the parameters of the problem. In particular the
dependence on the precision and the convergence properties of the solutions
of both the oscillator radius of the basis (the parameter unambiguously
fixing the square-integrable basis) and the form of the potential energy
contribution, will be treated in detail.

The convergence is crucial for obtaining stable approximations to the
solutions of problems expressed in terms of an (infinite) set of basis
functions, in our case an $L^2$ basis. This problem was repeatedly
investigated, mainly for bound-state solutions. As applications to
scattering problems, expressed in an $L^2$ basis, appeared, several
algorithms were suggested to accelerate the convergence of the results
within a restricted subset of the basis. For instance, Heller and Yamani 
\cite{kn:Heller2} used ``Kato correction'', and Revai et al.\ \cite{kn:revai}
introduced the ``Lanczos factor''. A more intuitive approximation was
proposed by Fillipov et al.\ \cite{kn:fil_rev2}.

The analysis of the AM equations presented in this work will be shown to
lead to (1) an optimal choice for the basis, given the potential, yielding
well-converging and stable solutions of the AM system, or, if the optimal
basis cannot be used, to (2) algorithmic procedures for solving the AM
system in an acceptable and controlled approximation. These algorithmic
procedures will depend on the specific choice made for the basis (i.e.\ the
oscillator radius chosen), but also on information on the asymptotic
behaviour of the expansion coefficients, as well as on the dynamical
equations themselves. In this way the algorithm used will depend on the
physical properties of the system considered.

In a forthcoming paper the different strategies for solving AM equations
will be applied explicitly to a number of problems from nuclear, atomic and
molecular physics.

\section{The AM in an Oscillator Basis}

Choosing an oscillator basis to describe some specific Hilbert subspace in
which to solve the Schr\"{o}dinger equation leads to the following form of
the latter in terms of a system of linear equations: 
\begin{equation}
\sum_{m=0}^\infty <n \mid \hat{H} - E \mid m> c_{m} = 0  \label{eq:AlgEqGen}
\end{equation}
where the coefficients $c_m$ are the expansion coefficients in the
oscillator basis of the wave function corresponding to the energy E 
\begin{equation}
\mid \Psi_E>=\sum_{n=0}^\infty c_{n} \ \mid n>,  \label{eq:PsiE}
\end{equation}
and subject to a typical boundary condition. For simplicity we have omitted
the angular momentum quantum number as well as the energy dependence of the
coefficients $c_n$.

The boundary conditions of quantum systems are traditionally expressed in
coordinate space, but can also be formulated in terms of the expansion
coefficients $c_n$. Indeed, for very large $n$, the dynamical equations are
reduced to a simple (in an oscillator basis a three-term recurrence) form
containing the kinetic energy operator solely. The equations can therefore
be solved analytically for very large $n$ \cite{kn:Yamani,kn:Fil,kn:Smirnov}%
. The above mentioned reduction of the equations only involves the
supposition that the potential energy matrix elements vanish for very large $%
n$, which has been shown to be an acceptable approximation for relatively
short-range interactions.

One obtains the following asymptotic behaviour for bound states: 
\begin{eqnarray}
c_{n}^{(as)} & \simeq & \sqrt{R_{n}} \ \exp(- \kappa R_{n}) / R_{n}
\label{eq:AlgCoefAsBound} \\
(\kappa & = & \sqrt{-2mE/\hbar^{2}})  \nonumber
\end{eqnarray}
and for continuum states: 
\begin{eqnarray}
c_{n}^{(as)} & \simeq & \sqrt{R_{n}} \ (j_{L}(k R_{n}) + \tan(\delta) \
n_{L}(k R_{n}))  \label{eq:AlgCoefAsCont} \\
(k & = & \sqrt{2mE/\hbar^{2}})   \nonumber
\end{eqnarray}
where $j_{L}$ and $n_{L}$ are the traditional Bessel and Neumann  special
functions.

A striking resemblance with the asymptotic forms of a wave function in
coordinate representation is observed, by replacing the radial coordinate in
the latter by the discrete value $R_n = \sqrt{4n+2L+3}$. A heuristic
argument for this observation is that $R_n$ corresponds to the turning point
of the oscillator in state $\mid n>$ with angular momentum $L$.

\section{A Simple Scheme for Solving the AM System of Equations}

In the previous section we introduced the infinite dimensional system of
linear equations obtained from the AM formulation, to be solved subject to a
proper boundary condition. As indicated earlier we will concentrate on the
scattering situation, and therefore only consider the asymptotic behaviour
of the $c_n$ corresponding to the continuum boundary condition.

A scheme for solving the linear system of equations quite naturally presents
itself. Under the assumption that the matrix elements potential energy %
\mbox{$<i \mid V \mid j>$} vanish for sufficiently large values of one of
the basis state indices $i$ or $j$, one chooses a limiting value $N$ to set
this vanishing point in terms of the basis states. In the region where the
potential matrix elements are neglected, the expansion coefficients $c_n$
are then given by (\ref{eq:AlgCoefAsCont}), and can be written as 
\begin{equation}
c_n = c_n^{(+)} + \tan(\delta) \ c_n^{(-)} \ \ \ (n \geq N)  \label{eq:zzzzz}
\end{equation}
assuming $N$ to be sufficiently large, with 
\begin{eqnarray}
c_{n}^{(+)} & = & \sqrt{R_{n}} \ j_{L}(k R_{n}) \ \ \ (n \geq N)  \nonumber
\\
c_{n}^{(-)} & = & \sqrt{R_{n}} \ n_{L}(k R_{n}) \ \ \ (n \geq N)
\label{eq:casytrad}
\end{eqnarray}

The choice for $N$ divides the linear system in three different regions:

\begin{itemize}
\item  a finite number of N equations with $0 < n < N$, in which the
potential energy matrix elements are fully taken into account. These
equations correspond to the ``internal region'' in terms of the basis states

\item  an infinite number of equations corresponding to $n > N$, in which
the potential energy matrix elements are neglected. These equations
correspond to the ``asymptotic region'', and are trivially fulfilled due to
the boundary condition

\item  the equation with $n = N$, in which the potential matrix elements are
neglected. This equation corresponds to the ``matching condition'', as it
couples the internal region through coefficient $c_{N-1}$ with the
asymptotic region through the phase-shift $\delta$
\end{itemize}

This scheme amounts to solving the following $N+1$ dimensional system of
linear equations for the $N$ coefficients $c_n$ with $n = 0, 1, .., N - 1$,
and the phase-shift $\delta$: 
\begin{eqnarray}
\left (
\begin{array}{cccc}
H_{00}-E & \cdots & H_{0,N-1} & 0 \\ 
\vdots & \vdots & \vdots & \vdots \\ 
H_{N-1,0} & \cdots & H_{N-1,N-1}-E & T_{N-1,N} \ c_{N}^{(-)} \\ 
0 & \cdots & T_{N,N-1} & {\cal T}_{N}^{(-)}
\end{array}
\right) \times \left (
\begin{array}{c}
c_{0} \\ 
\vdots \\ 
c_{N-1} \\ 
\tan(\delta)
\end{array}
\right)  \nonumber \\
= \left (
\begin{array}{c}
0 \\ 
\vdots \\ 
- T_{N-1,N} \ c_{N}^{(+)} \\ 
-{\cal T}_{N}^{(+)}
\end{array}
\right)  \label{eq:AlgEqStraight}
\end{eqnarray}
where 
\begin{eqnarray}
{\cal T}_{N}^{(+)} \, & = & \, (T_{N,N}-E) \ c_{N}^{(+)} + T_{N,N+1} \
c_{N+1}^{(+)}  \nonumber \\
{\cal T}_{N}^{(-)} \, & = & \, (T_{N,N}-E) \ c_{N}^{(-)} + T_{N,N+1} \
c_{N+1}^{(-)}
\end{eqnarray}
and $T$ stands for the kinetic energy operator.

As the potential matrix elements do not actually drop to zero exactly for $n
\geq N$, one should vary the value of $N$ to test the stability of the
solution. It turns out that this stability strongly depends on the specific
problem considered. As an example, the solution of the linear system of
equations for a nuclear two-cluster problem, in which the distance of the
clusters is considered as the degree of freedom, shows a rapid convergence
in terms of $N$. A solution for a monopole description of the nucleus, in
which the radius of the nucleus is considered to be the prominent degree of
freedom, shows a very slow convergence in terms of $N$. These results
indicate that one should be very careful when omitting potential energy
matrix elements, and that a proper study of the form of the equations is
necessary.

\section{An Analysis of the AM Equations}

\subsection{A Reformulation of the AM Equations}

To study the properties of the (in principle infinite dimensional) linear
system (\ref{eq:AlgEqGen}) to be solved, we will rewrite this set of
equations using the following substitution for the expansion coefficients $%
c_n$: 
\begin{equation}
c_n = c_{n}^{(as)} + c_{n}^{(0)}  \label{eq:cnDevAs}
\end{equation}
By this substitution, the coefficients $c_n$ are considered to represent a
deviation from the asymptotic behaviour, i.e.\ the coefficients $c_{n}^{(as)}
$. The coefficient $c_{n}^{(0)}$ then quantifies this deviation, which in
particular will be zero in the true asymptotic region (i.e.\ for very large $%
n$). The first term in (\ref{eq:cnDevAs}) is responsible for the long-range
behaviour of the system. The second term corresponds to the short-range
correction caused by the potential; in other words, in coordinate
representation the coefficients $\{c_n^{(0)}\}$ would represent that part of
the wave-function that is dominated by, and within the range of, the
potential. The knowledge of $c_{n}^{(0)}$ as a function of $n$ thus provides
a key element for determining a proper index $N$ distinguishing the internal
from the asymptotic region in terms of the basis functions.

Rewriting the original AM linear system of equations (\ref{eq:AlgEqGen}) in
the unknowns $(\{c_n\})$, yields an equivalent linear system in the unknowns 
$(\{c_{n}^{(0)}\}, \delta)$, where $\delta$ is defined by the following form
of the asymptotic expansion coefficients $c_{n}^{(as)}$: 
\begin{equation}
c_n^{(as)} = c_{n}^{(+)} + \tan (\delta) \ c_{n}^{(-)}  \label{eq:AsCoef}
\end{equation}

Because the asymptotic coefficients now appear for all $n$ in this
representation, they should be properly defined. In order to do so, we
consider the coordinate representation of the %ingoing and
outgoing asymptotic wave-functions $\Psi^{(+)}$ and $\Psi^{(-)}$, which are
originally defined as the two linearly independent solutions of 
\begin{equation}
(\hat{T} - E) \Psi = 0  \label{eq:PsiAsymp}
\end{equation}
where $\hat{T}$ is the kinetic energy operator. $\Psi^{(+)}$ is commonly
called the ``regular'' solution, and behaves properly for all $r$. $%
\Psi^{(-)}$ is the ``irregular'' solution, and has an irregular (infinite)
behaviour near the origin $r = 0$. To provide a regular character at the
origin for both $\Psi^{(+)}$ and $\Psi^{(-)}$, we redefine their equations
in the following way, as was suggested earlier by Heller and Yamani \cite
{kn:Heller1}: 
\begin{eqnarray}
(\hat{T} - E) \Psi^{(+)} & = & 0  \nonumber \\
(\hat{T} - E) \Psi^{(-)} & = & \beta_0 \ \Phi_{0}  \label{eq:PsiPlusPsiMinus}
\end{eqnarray}
The non-zero right-hand side in (\ref{eq:PsiPlusPsiMinus}), in which $%
\Phi_0(r) = <r \mid 0>$ represents the zero-quanta oscillator state, and
where $\beta_0$ equals 
\begin{equation}
\beta_0 = \hbar \omega \ \frac{\exp ( k^2/2 )}{k^{L+1}} \ \sqrt{\frac{2}{%
\Gamma(L+1/2)}} \ \frac{1}{\Gamma(-L+1/2)}  \label{eq:beta}
\end{equation}
accounts for a regular behaviour near the origin for the modified $\Psi^{(-)}
$: 
\begin{eqnarray}
\Psi^{(-)}(kr) \approx \left \{ 
\begin{array}{ll}
j_{L} (kr) & \mbox{for $ r \ll 1 $ } \\ 
n_{L} (kr) ) & \mbox{for $ r \gg 1 $}
\end{array}
\right.
\end{eqnarray}
In terms of a Fourier representation, using oscillator states as a basis,
one then has the following well-defined expansion: 
\begin{eqnarray}
\Psi^{(+)}(r) = \sum_{m = 0}^\infty <r \mid m> c_{m}^{(+)}  \nonumber \\
\Psi^{(-)}(r) = \sum_{m = 0}^\infty <r \mid m> c_{m}^{(-)}  \label{eq:eq144}
\end{eqnarray}
This provides a proper definition for the coefficients $c_{n}^{(+)}$ and $%
c_{n}^{(-)}$ in (\ref{eq:AsCoef}), the ``regular'', resp. ``irregular''
asymptotic coefficients, the explicit form of which can be found in \cite
{kn:Yamani,kn:Smirnov}. The equations for the asymptotic coefficients in
Fourier space are then: 
\begin{eqnarray}
\sum_{m=0}^\infty <n \mid \hat{T} - E \mid m> c_{m}^{(+)} & = & 0,  \nonumber
\\
\sum_{m=0}^\infty <n \mid \hat{T} - E \mid m> c_{m}^{(-)} & = & \beta_0 \
\delta_{n, 0}  \label{eq:AsCoefDef}
\end{eqnarray}
One notices an identical form for all equations, except for the single one
with $n = 0$.

Substituting the solutions of (\ref{eq:AsCoefDef}) for the asymptotic
coefficients in the set of equations (\ref{eq:AlgEqGen}), taking into
account the substitutions (\ref{eq:cnDevAs}) and (\ref{eq:AsCoef}), then
leads to the following linear system: 
\begin{eqnarray}
\sum_{m=0}^\infty <n \mid \hat{H} - E \mid m> c^{(0)}_{m} & + & \tan
(\delta) [\beta_0 \ \delta_{n, 0} + \sum_{m=0}^\infty <n \mid \hat{V} \mid
m> c_{m}^{(-)}] =  \nonumber \\
& - &\sum_{m=0}^\infty <n \mid \hat{V} \mid m> c_{m}^{(+)}
\label{eq:AlgEqNew}
\end{eqnarray}
Equation (\ref{eq:AlgEqNew}) now shows the influence of the potential energy
matrix elements on the behaviour of the system in a clear way.

Let us introduce the ``Dynamical Regular and Irregular Coefficients'' $%
V_n^{(+)}$ and $V_n^{(-)}$ as follows: 
\begin{eqnarray}
V_{n}^{(+)} & = & \sum_{m=0}^\infty <n \mid \hat{V} \mid m> c_{m}^{(+)} 
\nonumber \\
V_{n}^{(-)} & = & \sum_{m=0}^\infty <n \mid \hat{V} \mid m> c_{m}^{(- )}
\label{eq:SplusSminus}
\end{eqnarray}
or, in an alternative integral representation in three-dimensional
coordinate space: 
\begin{eqnarray}
V_{n}^{(+)} & = & \int_{0}^{\infty} dr \ r^{2} \ \Phi_{n}(r) \ \hat{V}(r) \
\Psi^{(+)} (kr)  \nonumber \\
V_{n}^{(-)} & = & \int_{0}^{\infty} dr \ r^{2} \ \Phi_{n}(r) \ \hat{V}(r) \
\Psi^{(-)} (kr)  \label{eq:SplusSminusInt}
\end{eqnarray}
where we use the coordinate representation of $\Phi_{n}(r)$, the n-quanta
oscillator function $<r \mid n>$: 
\begin{eqnarray}
\Phi_{n}(r) & = & (-1)^{n} \ N_{nL} \ \rho^{L} \ \exp(- \frac{1}{2}
\rho^{2}) \ L_{n}^{L+\frac{1}{2}}(\rho^{2}) \ \ \ \ \ (\rho = \frac{r}{b})
\label{eq:OscFun} \\
(N_{nL} & = & \sqrt{\frac{2 \ (n!)}{\Gamma(n+L+3/2)}} \frac{1}{b^{3/2}}\ ) 
\nonumber
\end{eqnarray}
Substitution of the Dynamical Coefficients (\ref{eq:SplusSminus}) in (\ref
{eq:AlgEqNew}) leads to 
\begin{equation}
\sum_{m=0}^\infty <n \mid \hat{H} - E \mid m> c^{(0)}_{m} + \tan (\delta) %
\left[ \beta_0 \ \delta_{n, 0} + V_{n}^{(-)} \right] = -V_{n}^{(+)}
\label{eq:AlgEqWithS}
\end{equation}

No approximations have been made so far to obtain this representation of the
AM dynamical equations. We have taken into account the main (asymptotic)
behaviour of the solutions in the equations, and used a regularization
scheme to achieve this in a well-defined way.

To reveal how the AM linear system can be solved in a numerically optimal,
or at least acceptable, way, an analysis of the Dynamical Coeeficients $%
V_n^{(+)}$ and $V_n^{(-)}$ seems imperative. It is indeed clear that, if
e.g.\ $V_n^{(+)}$ and $V_n^{(-)}$, from a given index $n$ on, are
sufficiently zero when compared to the other terms in (\ref{eq:AlgEqWithS}),
the equations are reduced in a controlled and secure way.

There is another important reason to investigate the behaviour of $%
V_{n}^{(+)}$. It is indeed well-known from the integral equation formulation
of quantum mechanics \cite{kn:Newton} that: 
\begin{equation}
\tan \delta_L \, = \, - \frac{m k}{\hbar^2} \int_{0}^{\infty} dr \,
\Psi^{(+)}(kr) \, \hat{V} (r) \, \Psi(k, r)  \label{eq:newton}
\end{equation}
where $\Psi(k, r)$ is the exact solution of the Schr\"{o}dinger equation
obtained with potential $\hat V$. In an oscillator representation, (\ref
{eq:newton}) reads as: 
\begin{equation}
\tan \delta_L \, = \, - \frac{m k}{\hbar^2} \sum_{n=0}^\infty V_n^{(+)} c_n
\end{equation}
from which we immediately recognize the importance of $V_n^{(+)}$ in the
convergence of the solutions of the continuous spectrum.

The analysis of $V_{n}^{(+)}$, $V_{n}^{(-)}$ must be done in terms of the
basis index $n$, of the oscillator parameter $b$ fixing the basis, and of
the potential energy parameters. Results emerging from such an analysis can
then be checked by studying the solutions $c_{n}^{(0)}$ and $\tan(\delta)$
as a function of the same parameters.

\subsection{A Model Analysis using the Gaussian Potential}

To analyze the behaviour of the AM set of linear equations and its defining
quantities in a more or less general way, we consider a simple gaussian
potential of the form 
\begin{equation}
\hat{V}(r) \ = \ V_0 \ \exp(- (\frac{r}{a})^2 )
\end{equation}
There are a number of reasons to justify such a choice: (1) an operator with
gaussian functional form is easily handled in a harmonic oscillator basis,
as matrix elements can be calculated using closed expressions, or simple
recurrence formulae; (2) (semi-) realistic potentials are often expressed
explicitly as a finite sum of gaussians, each with specific amplitude and
width parameters, and are of common use e.g.\ in nuclear physics
calculations; (3) a very large class of potentials can be expressed in terms
of a gaussian transform, such as e.g.\ a Yukawa or a Coulomb potential.

In general the matrix elements of the potential energy operator, due to the $%
r/a$ dependence of the latter, will depend on the ratio $b/a$. In the
specific case of a gaussian potential one has an explicit dependence on 
\begin{equation}
\gamma = (\frac{b}{a})^2  \label{eq:defgamma}
\end{equation}
This means that matrix elements of the gaussian potential energy operator in
an oscillator basis are invariant under all scale transformations of $a$ and 
$b$ which preserve $\gamma$, and different physical situations will lead to
the same matrix $<n\mid \hat{V} \mid m>$. In particular, small values of $%
\gamma$ are realised by small values of $b$ relative to $a$, or large values
of $a$ relative to $b$; this situation corresponds to a ``long range''
potential. Large values of $\gamma$ correspond likewise to a ``short range''
potential.

The potential is certainly not the only parametrized quantity characterizing
the set of AM equations. The oscillator radius of the oscillator basis is
another and equally important parameter, because of the repercussion on the
values of both the kinetic and potential matrix elements appearing in the
equations. In this section we will therefore study the behaviour of the AM
equations as a function of both the oscillator radius of the basis, and the
width of the gaussian potential. Actually, as indicated by (\ref{eq:defgamma}%
) only the ratio of these quantities is required, without any loss of
generality for the current analysis. We therefore make a specific choice of
potential parameters, namely $V_0 \ = \ -8 \ MeV$ and $a \ = \ 1 \ fm$, so
that the potential has both a discrete and continuous spectrum. The $\gamma$
ratio will then be varied by adapting the oscillator radius $b$.

A central theme in the analysis is the characterization for a rapidly
converging, and stabilized, solution of the AM equations. This is most
naturally done by searching for a maximal, limiting, value for the number of
basis states involved in a specific solution, depending on the problem
parameters used. As the asymptotic behaviour of the solution is governed
essentially by the kinetic energy term, one should therefore take into
account how both potential and kinetic matrix elements behave with respect
to one another. In other words, it is not sufficient to know about the
insignificance of potential matrix elements in absolute terms to decide
whether one has reached the asymptotic region, but rather consider some
relative insignificance with respect to the value of the kinetic matrix
elements.

When working in a coordinate representation, one can obtain a lot of
information concerning the wave-function by analyzing the Hamiltonian only,
and this both in terms of coordinate and energy ranges. The main reason for
this is that the Hamiltonian is globally defined in the whole coordinate
space. In a $L^2$ representation, in our case using an oscillator basis,
this is not such a straightforward matter. The main parameters determining
the rate of convergence, as well as the behaviour of the solution (the
wave-function in an oscillator representation), are of a ``global'' nature,
such as 
\mbox{$\sum_m <n
\mid \hat{V} \mid m> c_m \ = \ <n \mid \hat{V} \mid \Psi>$}, \mbox{$<n \mid
\hat{V} \mid \Psi^{(+)}>$} and so on. The matrix elements of the kinetic,
respectively potential energy operators are ``local'' quantities (cfr. the
value of these operators in a single point in coordinate space). The study
of these elements however reveals the peculiarities of the ``global''
quantities, and will help to understand their behaviour.

Concerning the kinetic energy term, two remarks are very important in the
context of an oscillator basis. The kinetic energy contribution has a very
simple representation in terms of the oscillator basis: (1) it is
proportional to the inverse square of the oscillator radius $b$, and (2) the
kinetic energy matrix has a tri-diagonal form, i.e.\ has non-zero matrix
elements only along the major diagonal and the first subdiagonals.

\subsubsection{Analysis of the Hamiltonian matrix elements}

\begin{figure}[tbp]
%\centerline{\psfig{figure=pot_3d.fig}} 
\centerline{\psfig{figure=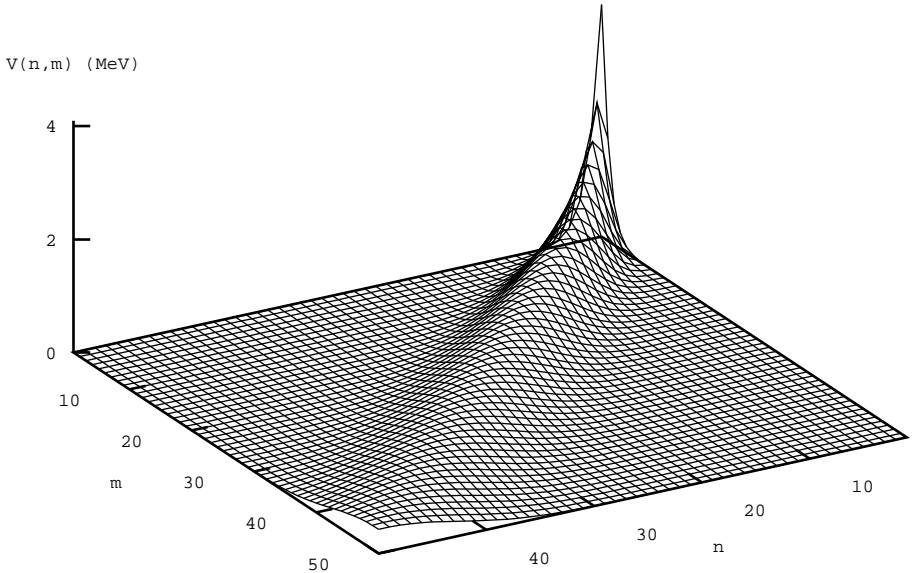}}
\caption{Matrix elements of a gaussian potential in an oscillator basis with
width parameter $b=0.75$. The horizontal axes are labelled by the basis
index, the vertical axis is in MeV.}
\label{fig:pot_3d}
\end{figure}

Fig. \ref{fig:pot_3d} displays the overall (qualitative) behaviour of a
typical gaussian potential matrix in some specific oscillator basis. The
main properties to be noted are the comportment of (1) the main diagonal
which falls off monotonically for large $n$, and of (2) the non-diagonal
matrix elements with non-zero values concentrated symmetrically around the
main diagonal. This structure of the potential contribution is certainly of
a promising nature, regarding the earlier remarks concerning the kinetic
energy contribution.

\begin{figure}[tbp]
%\centerline{\psfig{figure=pot_diag_abs.fig}} 
%\centerline{\psfig{figure=pot_diag_rel_b2.fig}} 
%\centerline{\psfig{figure=pot_diag_rel_0.fig}}
\centerline{\psfig{figure=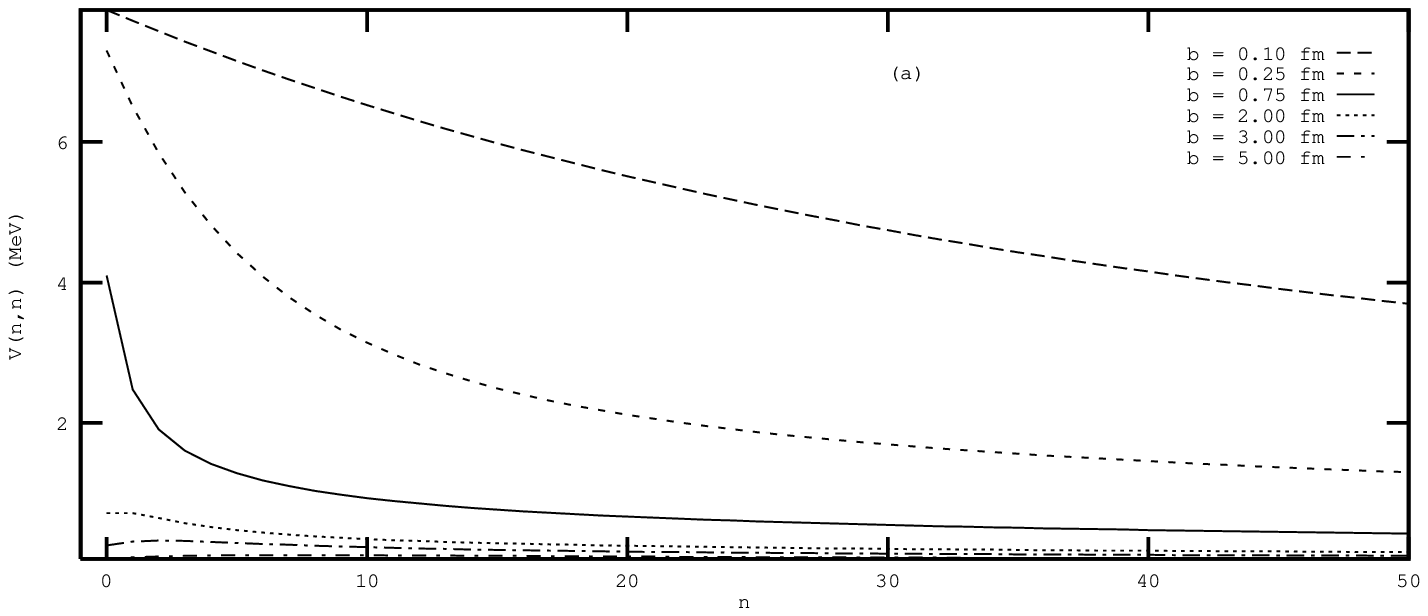}} \centerline{\psfig{figure=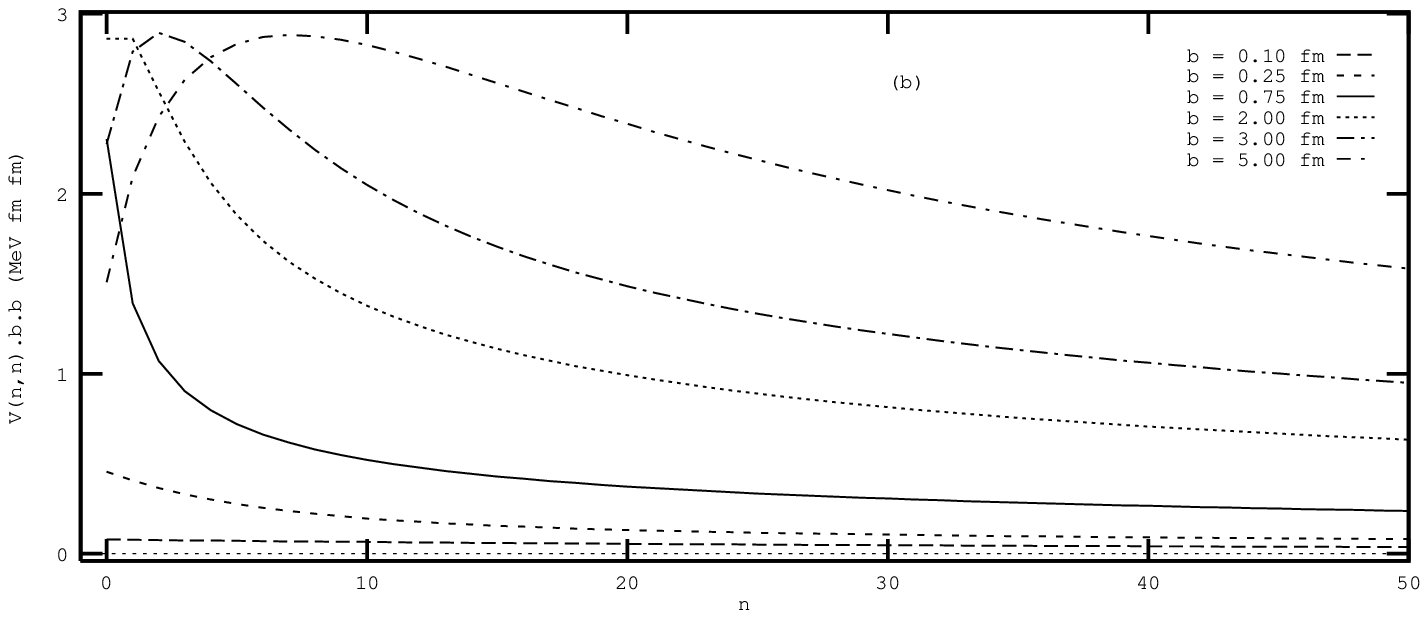}} %
\centerline{\psfig{figure=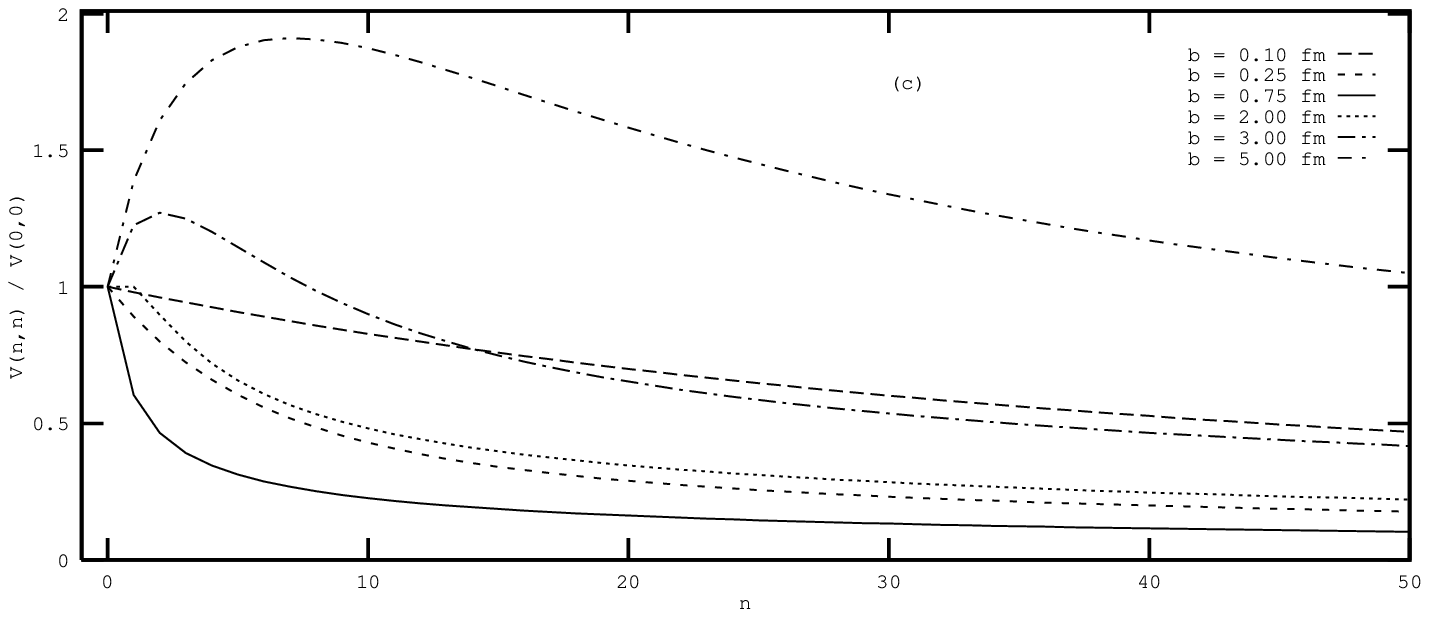}}
\caption{Diagonal matrix elements of a gaussian potential in oscillator
bases with different width parameter $b$: (a) exact values in $MeV$, (b)
values weighted with $b^2$ in $Mev \ fm^2$ and (c) values relative to $<0
\mid \hat{V} \mid 0>$. The horizontal axis labels the basis index.}
\label{fig:pot_diag}
\end{figure}

A more quantitative view of the diagonal behaviour of the potential matrix
is displayed in Fig. \ref{fig:pot_diag}, and this for a number of choices of
oscillator bases. Three different views are presented in this figure: (a)
the pure diagonal matrix elements; (b) the diagonal matrix elements
multiplied by the square of the oscillator radius (eliminating the kinetic
energy dependence on the choice of basis), and (c) the diagonal matrix
elements normalized with respect to the first one; both figures \ref
{fig:pot_diag} (a) and \ref{fig:pot_diag} (b) lead to Fig. \ref{fig:pot_diag}
(c) through this normalization. From these figures it should be clear that
one has to be careful when drawing conclusions for a proper choice of basis.
Fig. \ref{fig:pot_diag} (a) might suggest that very large values of the
oscillator radius $b$ are optimal. Fig. \ref{fig:pot_diag} (b) on the
contrary might suggest that very small values of $b$ are optimal. The
normalized Fig. \ref{fig:pot_diag} (c) finally suggests that, for the
current potential parameters, the interval for $b$ between $0.25 \ fm$ and $%
2.0 \ fm$ would be optimal. Fig. \ref{fig:pot_diag} at least reveals that
there are important differences in behaviour of the potential energy
contribution for different choices of the oscillator radius. Later on it
will be shown in a more quantitative way that the qualitative conclusion for
an optimal $b$ interval as suggested by Fig. \ref{fig:pot_diag} (c) is
correct. An interesting conclusion that is apparent from the figure is that
a (coordinate space defined) short-range potential turns out to have a
long-range character in an oscillator representation. As the oscillator
basis is used as a Fourier basis to portray the solution, this is a
well-known effect in terms of Fourier representation theory.

\begin{figure}[tbp]
%\centerline{\psfig{figure=pot_mid_abs.fig}}
%\centerline{\psfig{figure=pot_mid_rel_b2.fig}}
%\centerline{\psfig{figure=pot_mid_rel_0.fig}}
\centerline{\psfig{figure=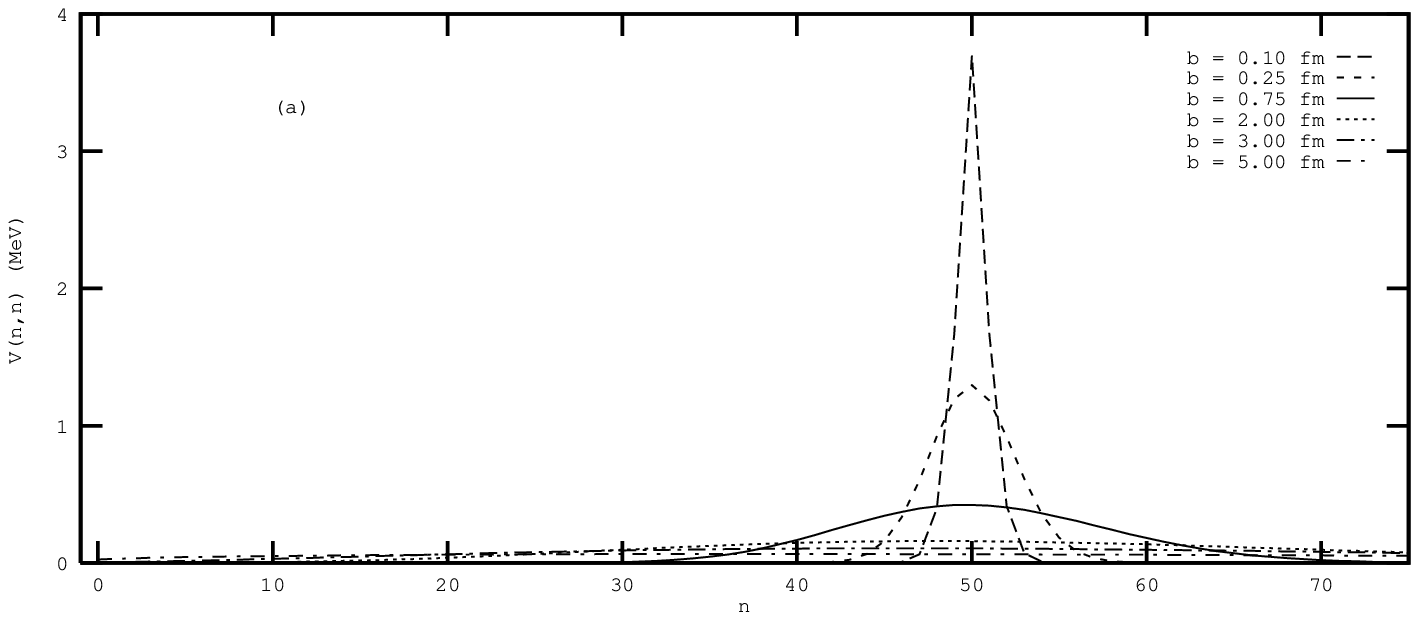}} \centerline{\psfig{figure=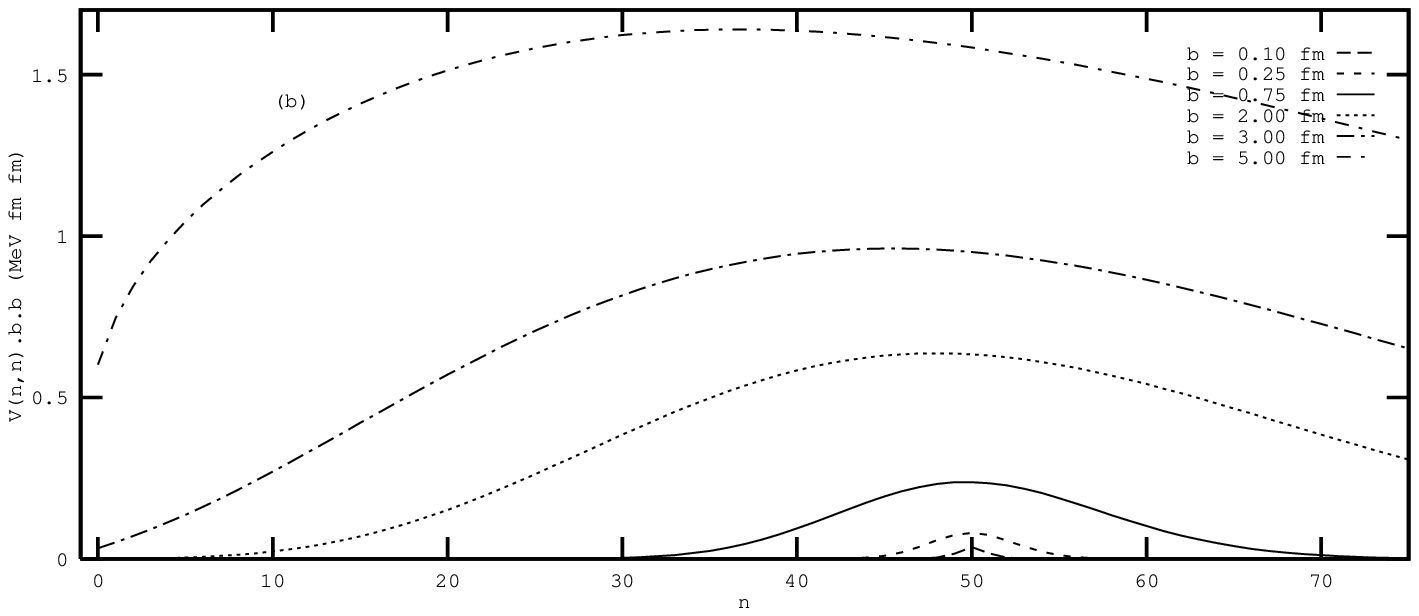}} %
\centerline{\psfig{figure=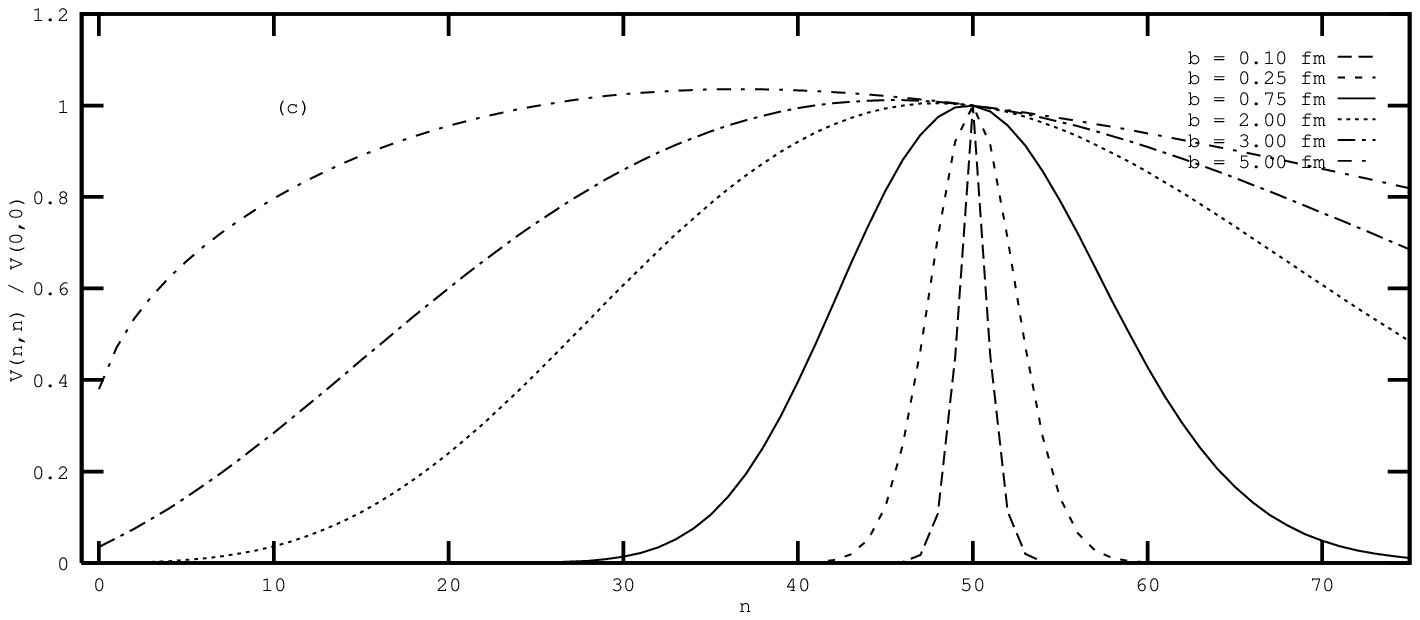}}
\caption{Matrix elements of a gaussian potential in oscillator bases with
different width parameter $b$ for row index $n=50$: (a) exact values in $MeV$%
, (b) values weighted with $b^2$ in $Mev \ fm^2$ and (c) values relative to $%
<50 \mid \hat{V} \mid 50>$. The horizontal axis labels the (column) basis
index.}
\label{fig:pot_mid}
\end{figure}

A quantitative view of the non-diagonal behaviour of the potential matrix
for the same bases displayed in Fig. \ref{fig:pot_diag} can be found in Fig. 
\ref{fig:pot_mid}, and this by showing a typical row for a fixed column
index ($n=50$ in this case). The same 3 views (pure, multiplied by $b^2$,
and normalized with respect to the diagonal matrix element with $n=50$) are
shown respectively in figures \ref{fig:pot_mid} (a), (b) and (c). Again a
strong dependence on the choice of the oscillator radius is remarked, but it
is much more difficult to draw conclusions for an optimally converging basis
from these figures. As this figure indicates that the potential energy
contribution is concentrated around the main diagonal of the matrix, but
with a relatively important distribution, it should already warn against
carelessly applying the simple but straightforward solution scheme presented
in section 2 of this paper!

The previous figures thus only allow us to initiate some general discussion,
but not to draw final and well-formed conclusions on the convergence
problem, let alone on the determination of an optimal basis and the number
of states involved in a stable solution.

\subsubsection{Analysis of the Dynamical Coefficients}

Up to now only local views of the potential contribution, i.e.\ matrix
elements and their behaviour, have been considered to provide possible hints
for a properly converging solution. The new representation of the AM
equations presented earlier in this section introduced some new, global
quantities which might be of interest in the convergence analysis. These are
the Dynamical Coefficients $V_n^{(+)}$ and $V_n^{(-)}$, and they combine all
matrix elements of row $n$ as indicated above. If, and when, these
quantities would be (sufficiently) zero from a given value $N$ on, this
would certainly determine the maximal number of oscillator states to be
considered for a converging solution. As $V_n^{(+)}$ and $V_n^{(-)}$ depend
on both the potential and the basis parameters, they will be analyzed as a
function of the ratio of potential width to oscillator radius. As $V_n^{(+)}$
and $V_n^{(-)}$ also depend explicitly on energy, this will be an additional
parameter to consider.

\begin{figure}[tbp]
%\centerline{\psfig{figure=s_plus_3d.fig}} 
%\centerline{\psfig{figure=s_minus_3d.fig}} 
\centerline{\psfig{figure=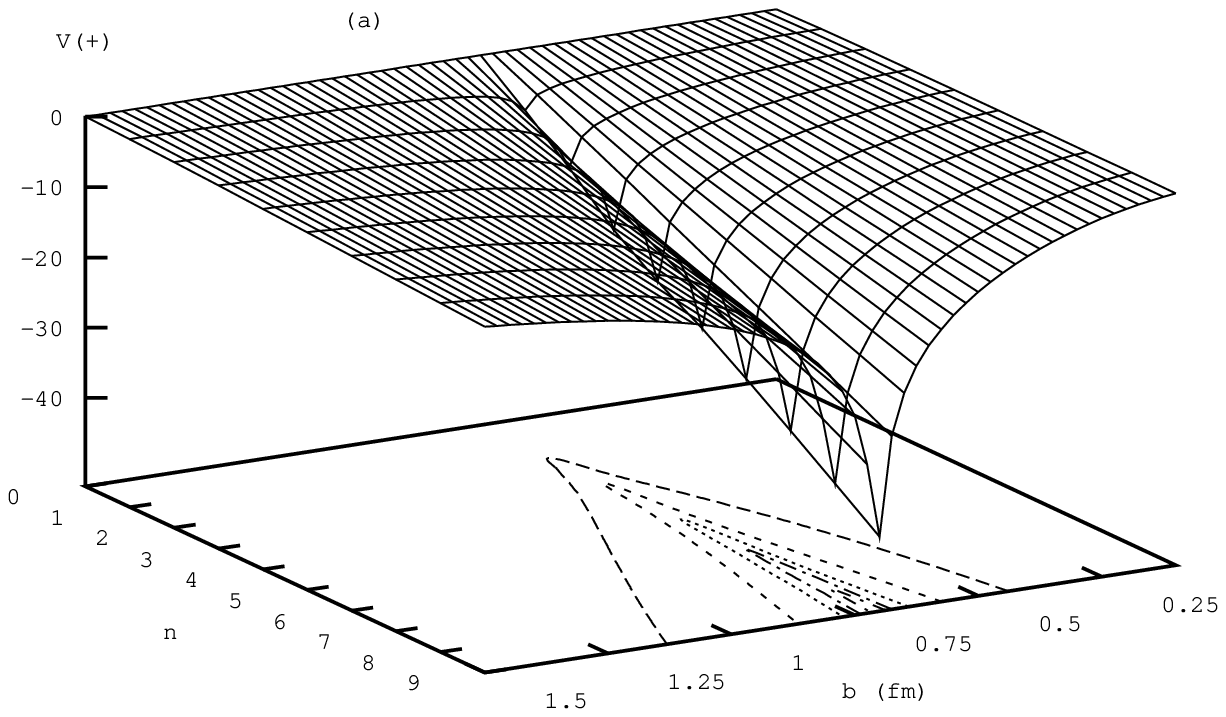}} \centerline{\psfig{figure=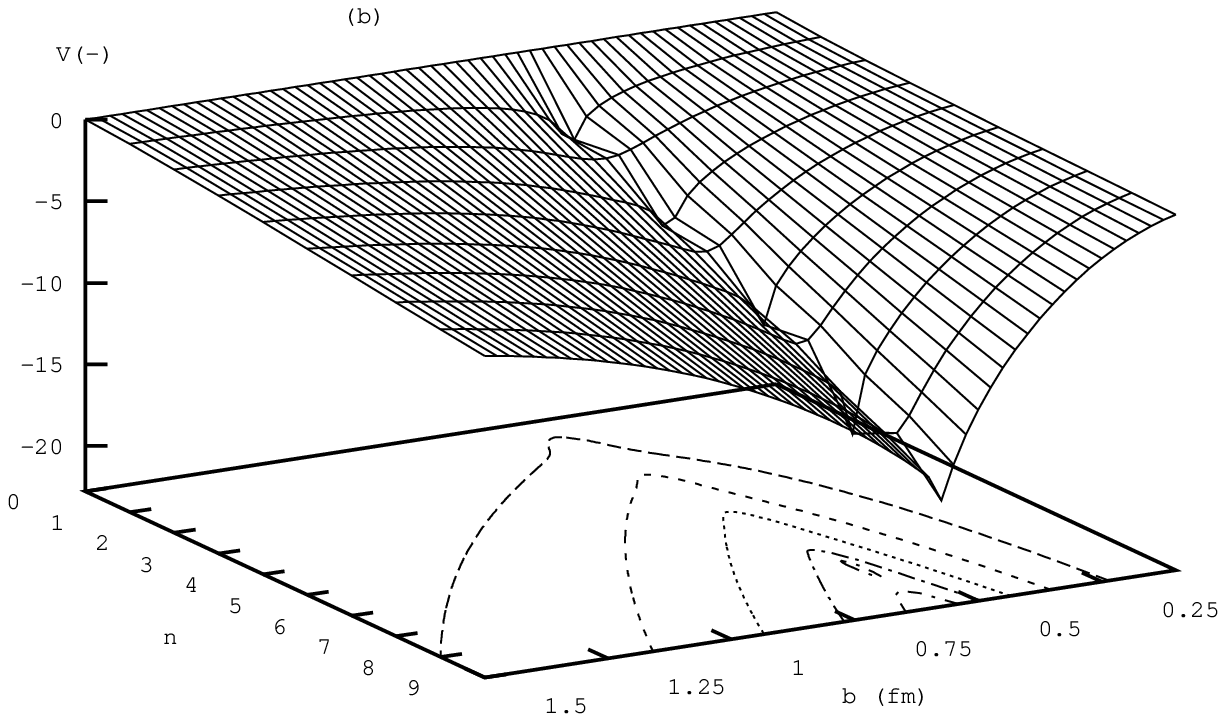}}
\caption{The Dynamical Coefficients $\left| V_n^{(+)} / V_0^{(+)} \right|$
(figure a, top) and $\left| V_n^{(-)} / V_0^{(-)} \right|$ (figure b,
bottom) as a function of $b$ and $n$, for an energy of $1 \ MeV$. The
vertical axis is logarithmic.}
\label{fig:s_3d}
\end{figure}

A qualitative view of $V_n^{(+)}$ is shown in Fig. \ref{fig:s_3d} (a) as a
function of both the index $n$ and the ratio $b/a$; the quantity was
normalized with respect to $V_0^{(+)}$, and the scale is a logarithmic one.
From this figure one immediately notices that a well-defined value of the
ratio of the potential to oscillator radius zeroes the quantities for
already very small values of $n$. Fig. \ref{fig:s_3d} (b) depicts the same
view for $V_n^{(-)}$, carrying an analogous conclusion. Although both
optimal values do not coincide completely, they are sufficiently close to
define a narrow interval of optimal values for a swiftly converging
solution. Indeed, if we consider a value $N$ for which $\mid V_N^{(+)} /
V_0^{(+)} \mid \ < \epsilon$ and $\mid V_N^{(-)} / V_0^{(-)} \mid\ < \epsilon
$ (with e.g.\ $\epsilon < 10^{-6}$), we can assume to have reached the
asymptotic region. The value $N$ then determines the number of basis
functions needed to obtain a well-converged solution. Indeed, the solutions
of (\ref{eq:AlgEqNew}) are deviations with respect to the (known) asymptotic
solutions, and become zero when both $V_n^{(+)}$ and $V_n^{(-)}$ are
(sufficiently) zero. In other words, the value $N$ from which on $V_n^{(+)}$
and $V_n^{(-)}$, and thus the solutions $c_n^{(0)}$, are zero determines
what would be called in RGM terminology the matching point between the
internal and the asymptotic region. It is important to remark that $N$ is
determined prior to solving the AM equations, and is obtained from the
simple knowledge of the potential energy matrix elements.

\begin{figure}[tbp]
%\centerline{\psfig{figure=s_plus_vs_b.fig}}
%\centerline{\psfig{figure=s_minus_vs_b.fig}}
\centerline{\psfig{figure=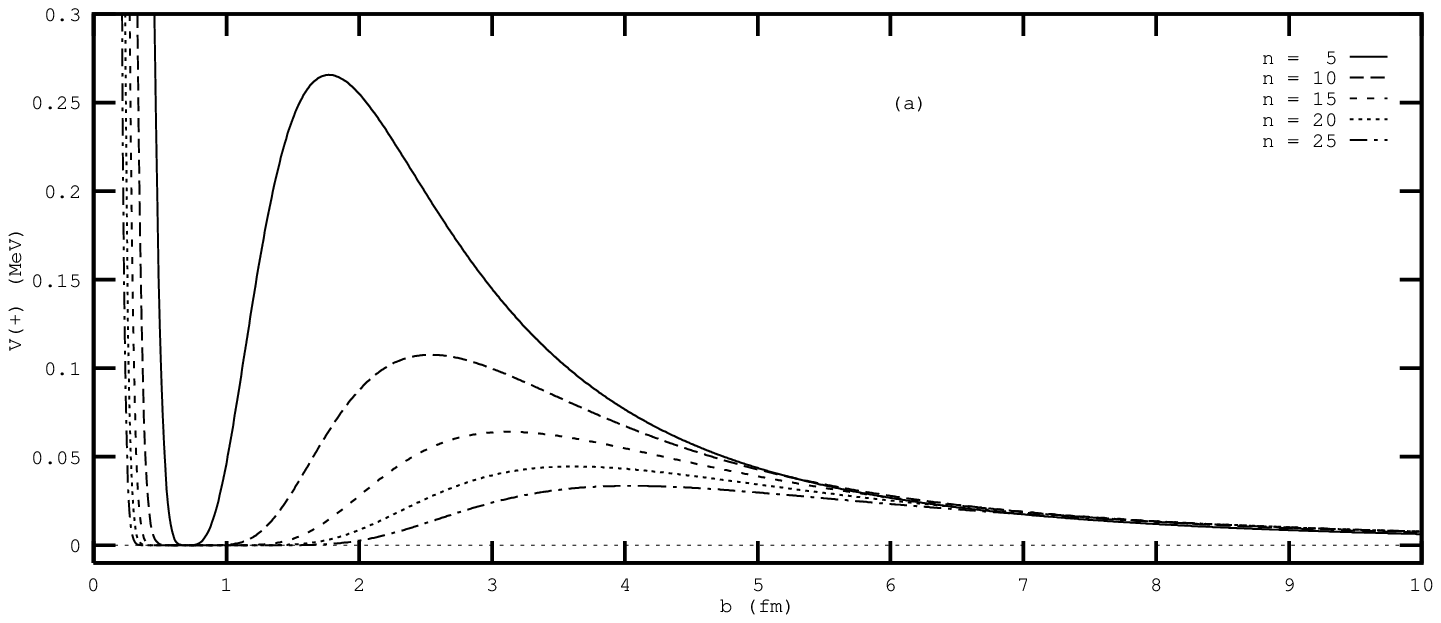}} \centerline{\psfig{figure=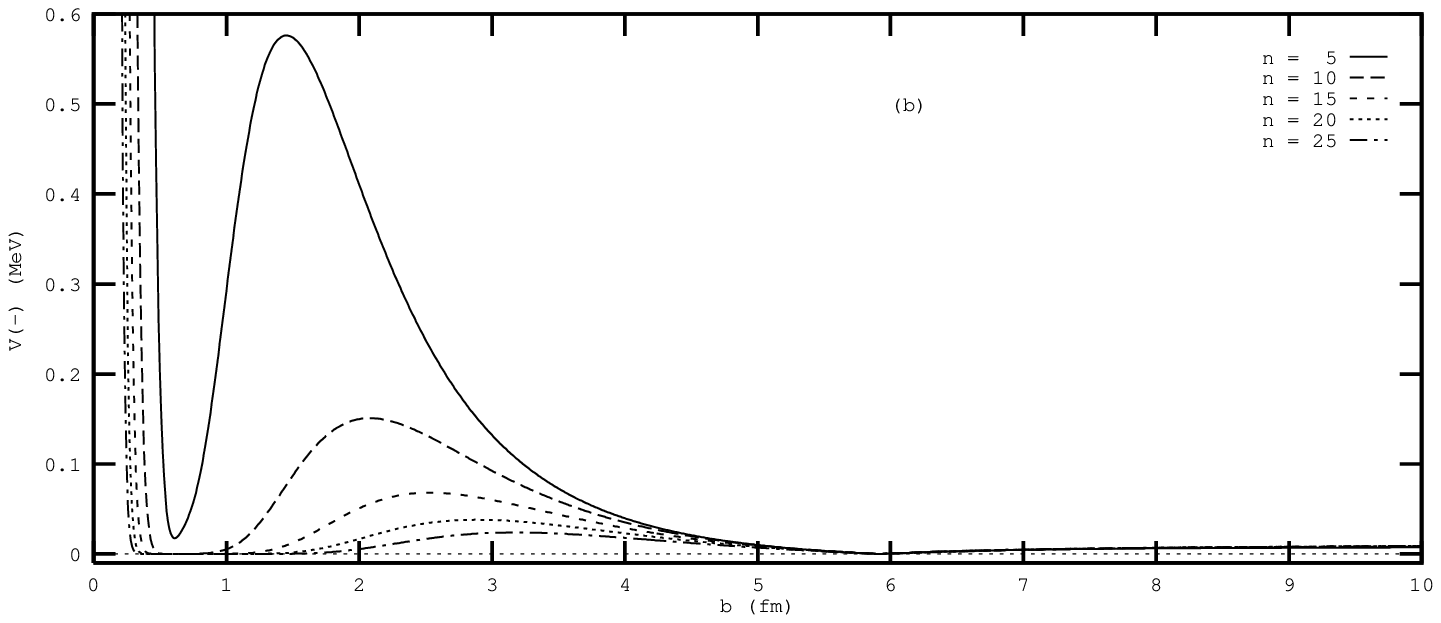}}
\caption{$V_n^{(+)}$ (figure a, top) and $V_n^{(-)}$ (figure b, bottom) as a
function of $b$ for various values of $n$ and an energy of $1 \ MeV$.}
\label{fig:s_2d}
\end{figure}

To corroborate the fact for an optimal choice of basis for a given gaussian
potential, a quantitative view of $V_n^{(+)}$ and $V_n^{(-)}$ as a function
of the radius parameter $b$ is shown in figures \ref{fig:s_2d} (a) and (b)
for a selected number of $n$ values. From this figure one can determine for
which values of the radius parameter one obtains a properly converged
solution, by considering the intersection of both optimal intervals for $%
V_n^{(+)}$ and $V_n^{(-)}$. One notices that for a potential consisting of a
single gaussian term, a very limited number of basis states is necessary for
a proper convergence. For more intricate potential forms this will not
necessarily be the case, although an optimal value for the radius parameter,
be it associated with a relatively larger value of $N$, will still be
available.

\begin{figure}[tbp]
%\centerline{\psfig{figure=s_plus_3d_vs_e.fig}} 
%\centerline{\psfig{figure=s_minus_3d_vs_e.fig}} 
\centerline{\psfig{figure=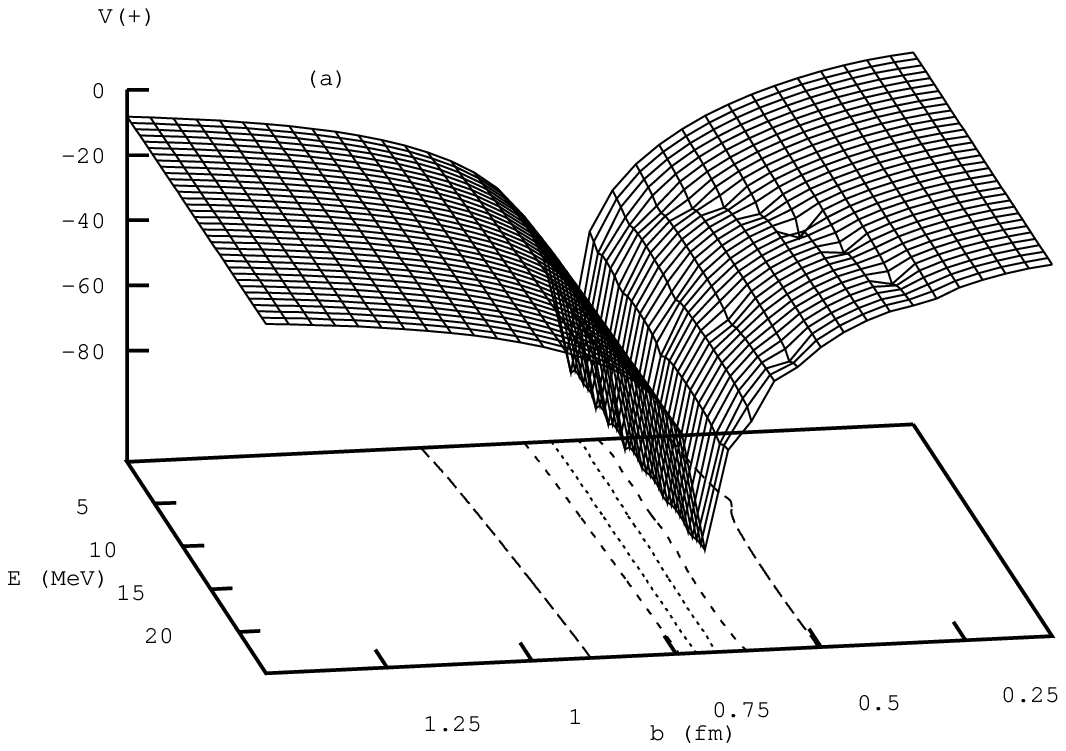}} \centerline{\psfig{figure=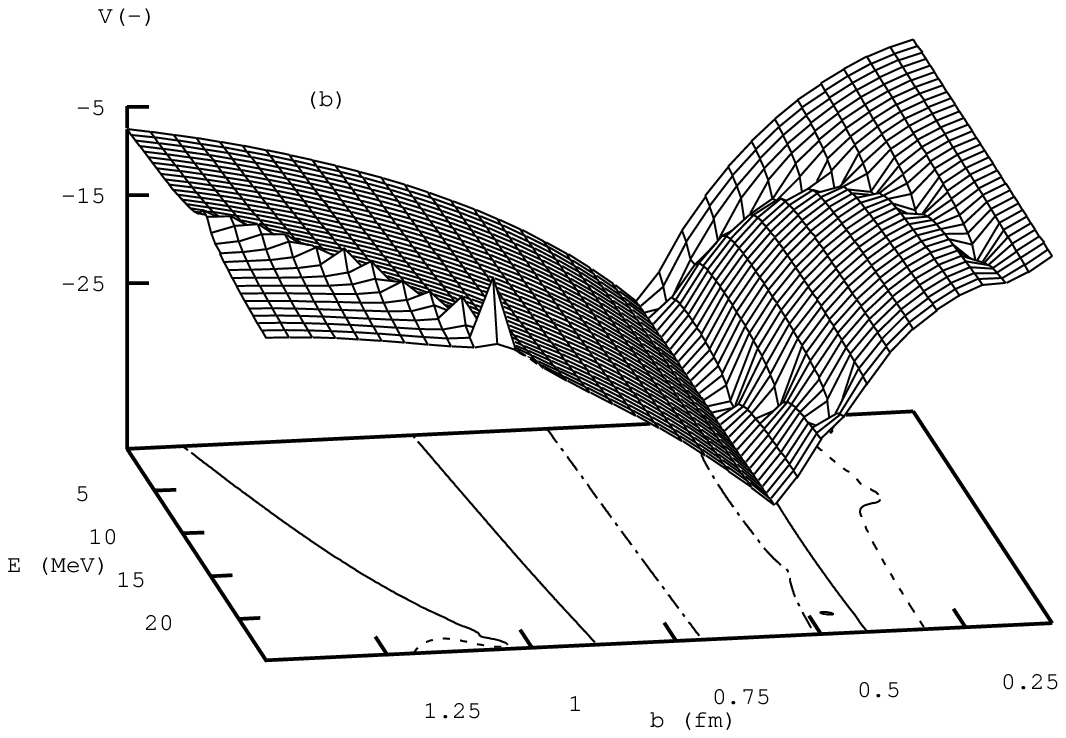}}
\caption{$\left| V_n^{(+)} / V_0^{(+)} \right|$ (figure a, top) and $\left|
V_n^{(-)} / V_0^{(-)} \right|$ (figure b, bottom) as a function of $b$ and $E
$, for $n=20$. $E$ values are in $MeV$ and the vertical axis is logarithmic.}
\label{fig:s_3d_vs_e}
\end{figure}

To check the dependence of the so-called ``optimal'' $b$ values (or regions)
as a function of energy, Fig. \ref{fig:s_3d_vs_e} displays the values of
both $V_n^{(+)}$ and $V_n^{(-)}$, relative to respectively $V_0^{(+)}$ and $%
V_0^{(-)}$ for varying $b$ and energy for a fixed $n$. It is important to
notice that, at least for the gaussian potential, the optimal value for the
oscillator radius is independent of energy.

\begin{figure}[tbp]
%\centerline{\psfig{figure=s_plus_vs_n.fig}} 
%\centerline{\psfig{figure=s_minus_vs_n.fig}} 
\centerline{\psfig{figure=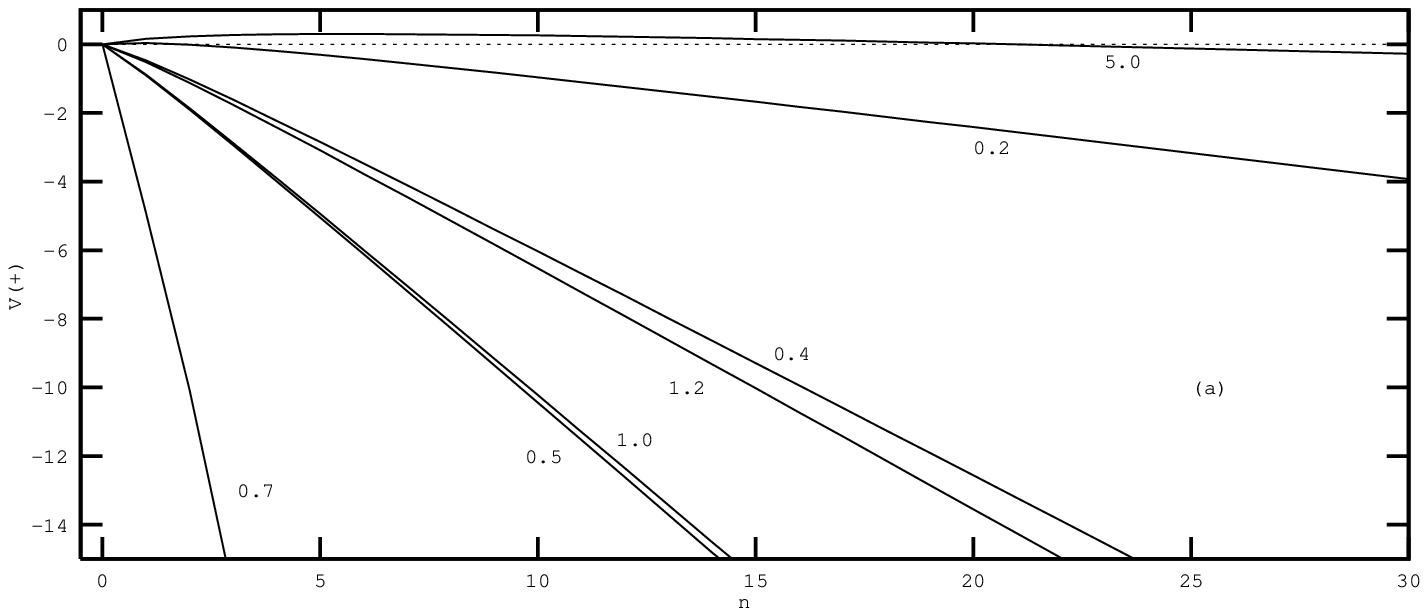}} \centerline{\psfig{figure=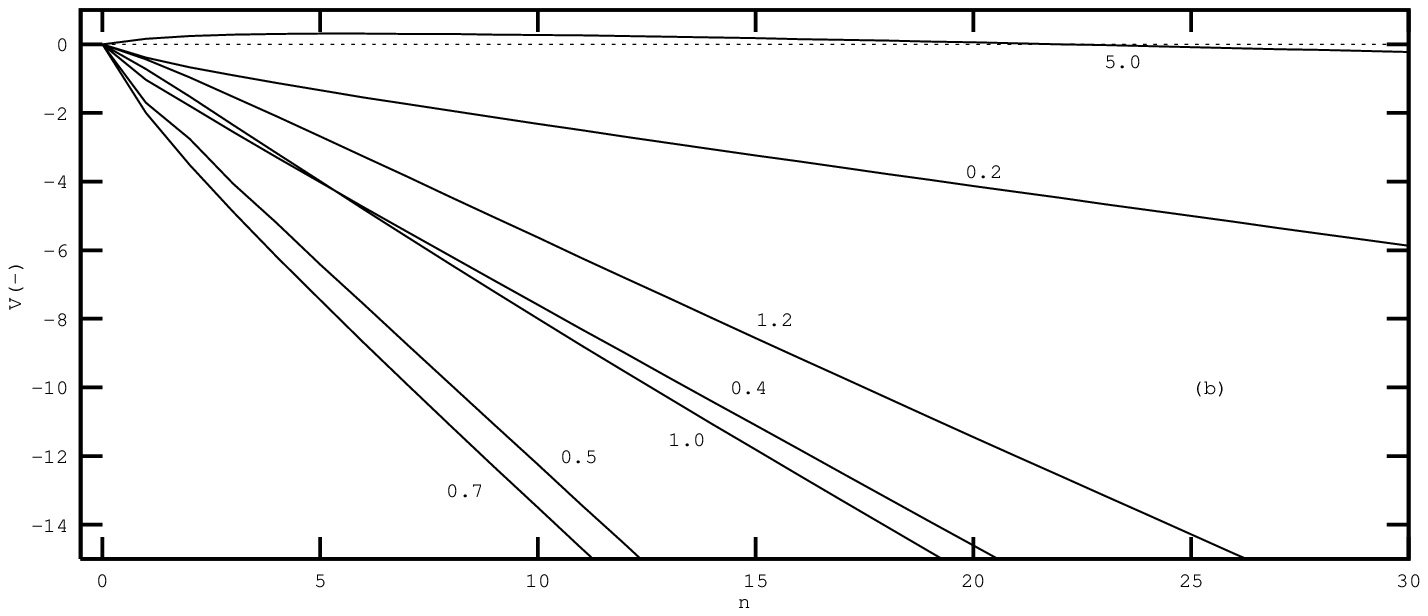}}
\caption{Logarithmic plot of $\left| V_n^{(+)} / V_0^{(+)} \right|$ (figure
a, top) and $\left| V_n^{(-)} / V_0^{(- )} \right|$ (figure b, bottom) as a
function of $n$ for several values of the oscillator radius $b$ and an
energy of $1 \ MeV$. The vertical axis shows orders of magnitude.}
\label{fig:s_vs_n}
\end{figure}

Finally Fig. \ref{fig:s_vs_n} shows the quantitative behaviour of $V_n^{(+)}$
and $V_n^{(-)}$, again relative to respectively $V_0^{(+)}$ and $V_0^{(-)}$,
as a function of $n$. From these figures one easily obtains an optimal
number of basis functions, given the specific parameters of the problem and
the precision considered. As a hands on example, Fig. \ref{fig:s_vs_n}
indicates that, for a precision of about $10^{-8}$, less than 12 basis
functions are needed for all $b$ values between $0.5 \ fm$ and $1 \ fm$.

The conclusions above indicate the importance of the $V_n^{(+)}$ and $%
V_n^{(-)}$ quantities, and a closer investigation imposes itself. A closed
expression for $V_n^{(+)}$ can be obtained in a straightforward way as: 
\begin{eqnarray}
V_{n}^{(+)} = V_{0} \frac{(1-2\gamma)^{n}}{(1+2\gamma)^{n+L+3/2}} k^{L}
\exp(-\frac{1}{2}k^{2}\frac{1}{1+2 \gamma}) N_{nL} L_{n}^{L+1/2}(\frac{k^{2}%
}{1-4 \gamma^{2}})  \label{eq:V_plus}
\end{eqnarray}
for which the full calculation is reproduced in appendix A.

This expression features a single minimum in terms of $\gamma$, namely for $%
\gamma = 1/2$ or $b=a/\sqrt{2}$, with a value 
\begin{equation}
V_{n}^{(+)} = V_{0} N_{nL} \frac{1}{4} \frac{1}{n!} (\frac{k}{2})^{2n+L+1/2}
\exp(-\frac{1}{4} k^{2})
\end{equation}
which drops off to zero very fast in terms of $n$ because of its following
behaviour: 
\begin{equation}
V_{n}^{(+)} \approx \frac{1}{n! \, R_{n}^{L+1/2}}
\end{equation}

Both position and behaviour of $V_n^{(+)}$ around this minimum were already
apparent from the foregoing figures.

The asymptotic behaviour of $V_n^{(+)}$ for both small and large values of $%
\gamma$ is also readily obtained. For small values of $\gamma$ (i.e.\ small
values of the oscillator radius $b$ compared to the potential width $a$)
equation (\ref{eq:V_plus}) yields the following asymptotic form, valid for
large $n$: 
\begin{eqnarray}
V_{n}^{(+)} & \approx & V_{0} \, \exp \{ -\gamma R_{n}^{2} \} \, c_{n}^{(+)}
\nonumber \\
& \approx & V_{0} \, \exp \{ -\gamma R_{n}^{2} \} \, \sqrt{2 R_{n}} \, j_{L}
(k b R_{n})  \label{eq:V_plus_asym_small_b}
\end{eqnarray}
This expression consists of two factors in coordinate representation, the
gaussian potential $\hat{V}(r)$ and the (asymptotic) wave function $%
\Psi^{(+)}(k, r)$, both evaluated in one and the same discrete point $r =
bR_n$ ($R_n$ is the classical turning point, cfr.\ (\ref{eq:AlgCoefAsBound})
and (\ref{eq:AlgCoefAsCont})). For $\gamma$ approaching zero the expression
shows a slowly decreasing behaviour of $V_{n}^{(+)}$ as a function of $n$.

For large values of $\gamma$ one obtains the following asymptotic form valid
for large $n$ 
\begin{eqnarray}
V_{n}^{(+)} & \approx (-1)^{n} & V_{0} \, \frac{1}{2 \gamma} \sqrt{\frac{2}{k%
}} \exp \left \{ - \frac{ R_{n}^{2} + k^{2}}{4\gamma} \right \}\, I_{L+1/2} (%
\frac{k}{2\gamma} R_{n} )  \label{eq:Spluslargeb}
\end{eqnarray}
which for the limiting case of $\gamma$ now approaching infinity again
displays a slowly decreasing behaviour as a function of $n$.

Both very large and very small values of $\gamma$ will thus lead to rather
badly converging solutions, which essentially means that a large number of
basis functions will have to be considered when $\gamma$ reaches towards
limiting values.

This discussion on asymptotic behaviour confirms conclusions already
apparent from the numerically calculated figures of $V_n^{(+)}$.

\begin{figure}[tbp]
\centerline{\psfig{figure=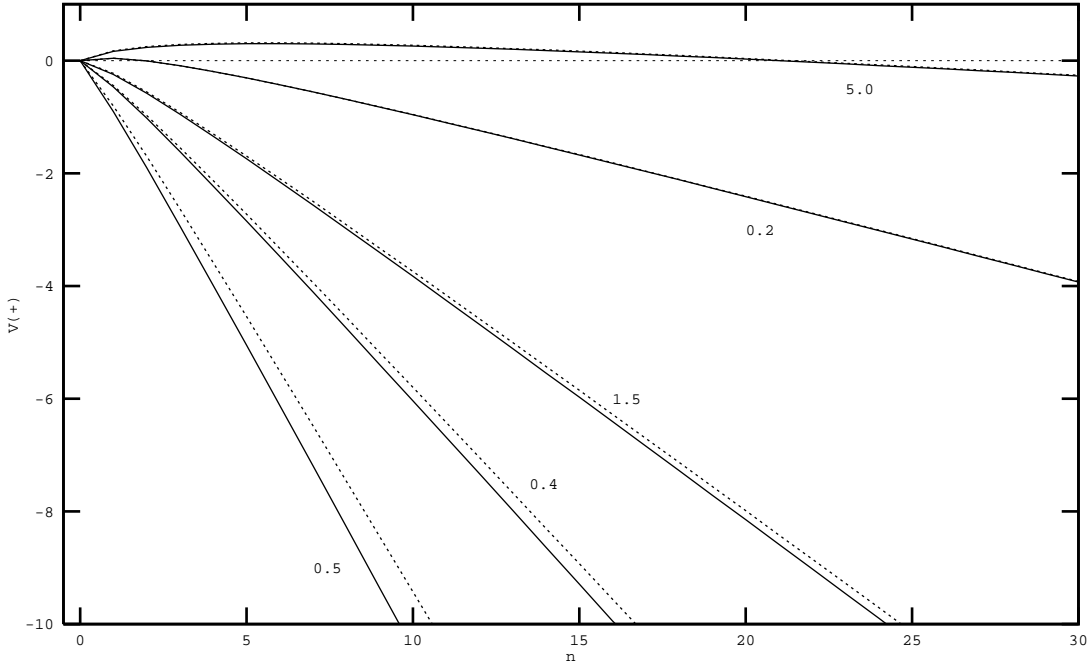}}
\caption{$\left| V_n^{(+)} / V_0^{(+)} \right|$ compared to the asymptotic
formulae discussed in the text, as a function of $n$ in a logarithmic plot.
The vertical axis shows orders of magnitude }
\label{fig:s_plus_vs_n_asy}
\end{figure}

Fig. \ref{fig:s_plus_vs_n_asy} indicates that the asymptotic formulae are
quite valid for a broad range of $b$ values, and as such, can often be used
to obtain reasonable values for $N$ for properly converged solutions.
Indeed, as the main factor which defines the decreasing behaviour of $%
V_n^{(+)}$ for small, respectively large values of $\gamma$, is the
potential term, which has the value $\exp\{-\gamma R_n^2\}$, respectively $%
\exp\{-\frac{R_n^2}{4 \gamma}\}$, as can be seen in equations (\ref
{eq:V_plus_asym_small_b}) and (\ref{eq:Spluslargeb}). So, if one considers
e.g.\ $10^{-6}$ to be a reasonable measure for precision, one can write
approximately 
\begin{eqnarray}
\left| \frac{V_n^{(+)}}{V_0^{(+)}} \right| \ & \approx & \ \exp\{-\gamma
R_n^2\} = 10^{-6} \ \ \makebox[1.5 cm]{(small} \gamma)  \nonumber \\
\left| \frac{V_n^{(+)}}{V_0^{(+)}} \right| \ & \approx \ & \exp\{-\frac{R_n^2%
}{4 \gamma}\} = 10^{-6} \ \ \makebox[1.5 cm]{(large} \gamma)
\label{eq:upperlimN}
\end{eqnarray}
which immediately leads to the respective values $N \approx 14 / (4 \
\gamma) $ (for small $\gamma$) and $N \approx 14 \gamma)$ (for large $\gamma$%
); for a value of $b/a = 0.2$ one obtains $N \approx 90$, and for $b/a = 3.0$
$N \approx 125$.

The asymptotic behaviour of $V_N^{(+)}$ can be used to an even better
extent, by taking it into account when solving the AM system of equations.
This will be pursued in full detail in a forthcoming section.

Although more intricate to develop, an analogous analysis can be made for $%
V_n^{(-)}$, leading to corresponding conclusions.

\subsubsection{Analysis of the phase shifts}

\begin{figure}[tbp]
\centerline{\psfig{figure=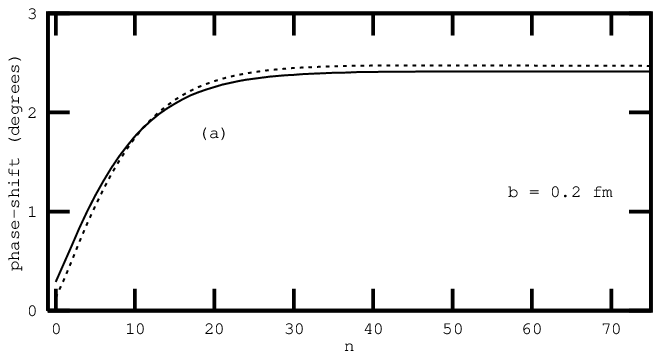} 
 \psfig{figure=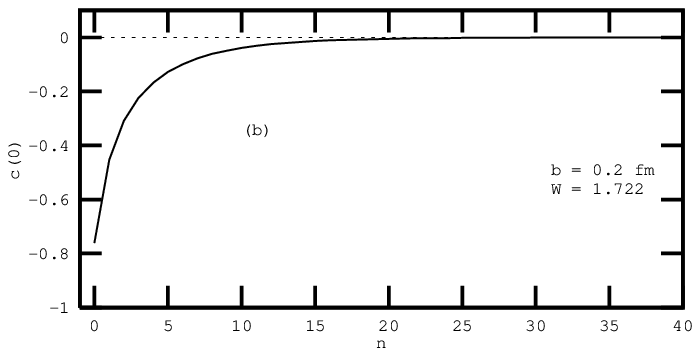}} 
\centerline{\psfig{figure=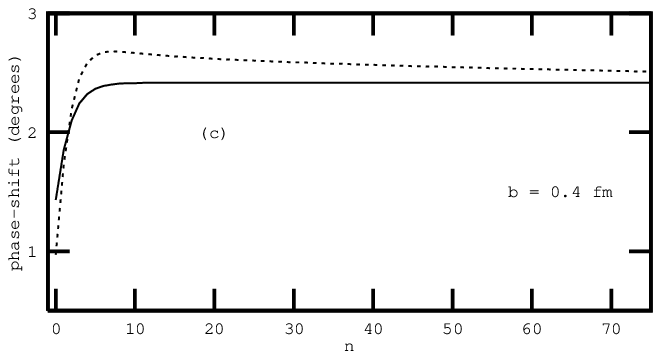} 
 \psfig{figure=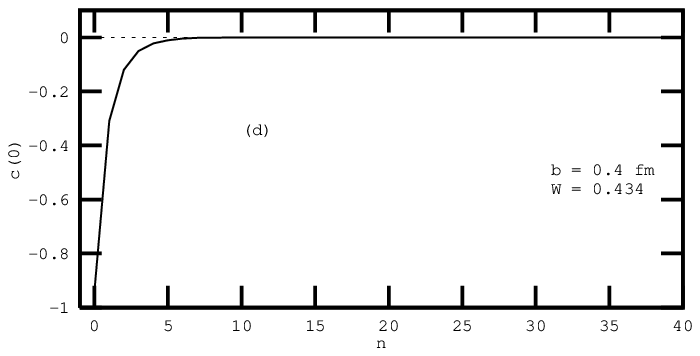}} 
\centerline{\psfig{figure=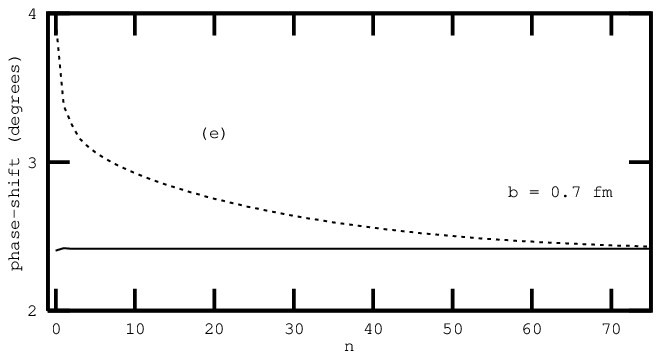} 
 \psfig{figure=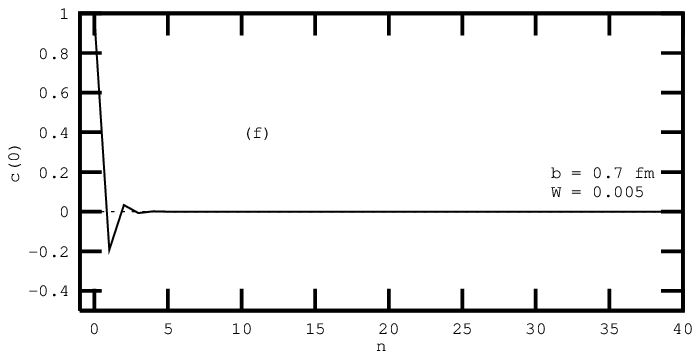}} 
\centerline{\psfig{figure=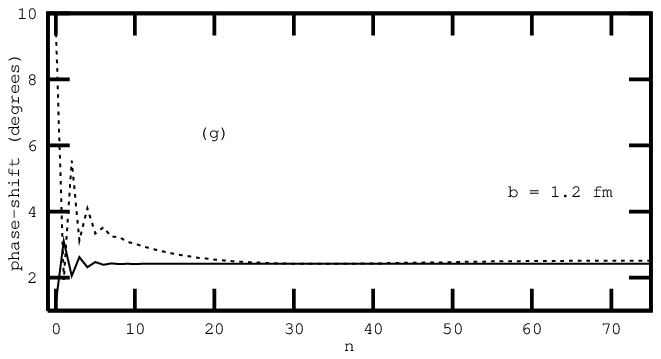} 
 \psfig{figure=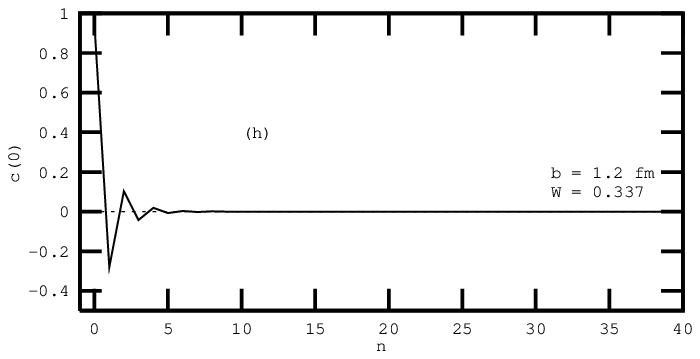}} 
\centerline{\psfig{figure=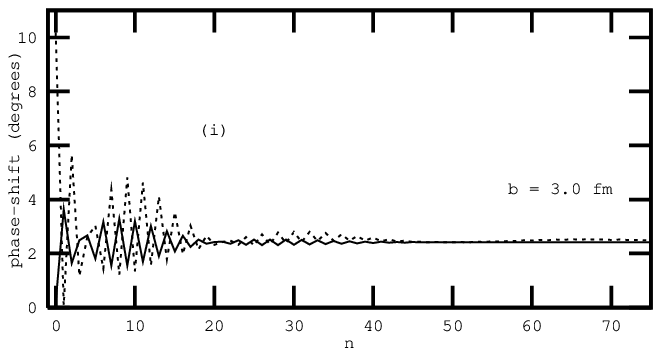} 
 \psfig{figure=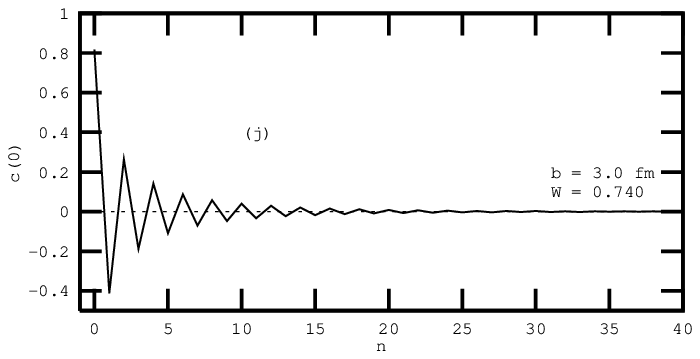}}
\caption{ Left side figures: phase shift $\protect\delta$ (degrees) as a
function of the number of basis states involved in the calculations; the
energy was fixed at $1 \ MeV$. Dotted lines refer to the solutions obtained
in the original (``simple'') formulation, full lines refer to solutions
obtained in the reformulated version. Right side figures: normalized $%
c_n^{(0)}$ values (see text) for a calculation with 100 basis states. The
norm $W$ is given as a percentage.}
\label{fig:phas_vect_vs_n}
\end{figure}

In Fig. \ref{fig:phas_vect_vs_n} we show the phase shifts obtained at an
energy of $E=1 \ MeV$ for different values of $\gamma$ (actually in all
further results $a = 1.0 \ fm$, so that $b$ and $\gamma$ coincide), as a
function of the number of basis states involved in the calculation. One
notices an important gain in convergence speed (and thus precision) when
using the reformulated version of the AM equations (\ref{eq:AlgEqWithS}),
although small and large $b$ (or $\gamma$) values still require an important
number of basis states.

On the same figure the normalized $c_n^{(0)}$ solutions are shown for
different values of $b$ ($\gamma$), for a calculation involving 100 basis
states. These results are normalized with respect to the norm $W$ 
\begin{equation}
W \ = \ \frac{\sum_{n=0}^{99} \mid c_n^{(0)} \mid ^2} {\sum_{n=0}^{99} \mid
c_n \mid ^2}
\end{equation}
which is also include in the figures. One notices from this value that in
the ``optimal $b$ ($\gamma$)'' region very few, very small $c_n^{(0)}$
contribute to the solution.

The results displayed on Fig. \ref{fig:phas_vect_vs_n} confirm our
suggestion made by analysing the behaviour of $V^{(+)}$ and $V^{(-)}$, i.e.
that for optimal values of $b$ ($\gamma$) convergence is achieved with less
than 12 basis functions. They also confirm that for $b=0.2$ one can use
approximately 75, and for $b=3.0$ approximately 100 basis states for
reasonable convergence; the latter were indeed overestimated by the numbers
obtained from equation (\ref{eq:upperlimN}).

\begin{figure}[tbp]
\centerline{\psfig{figure=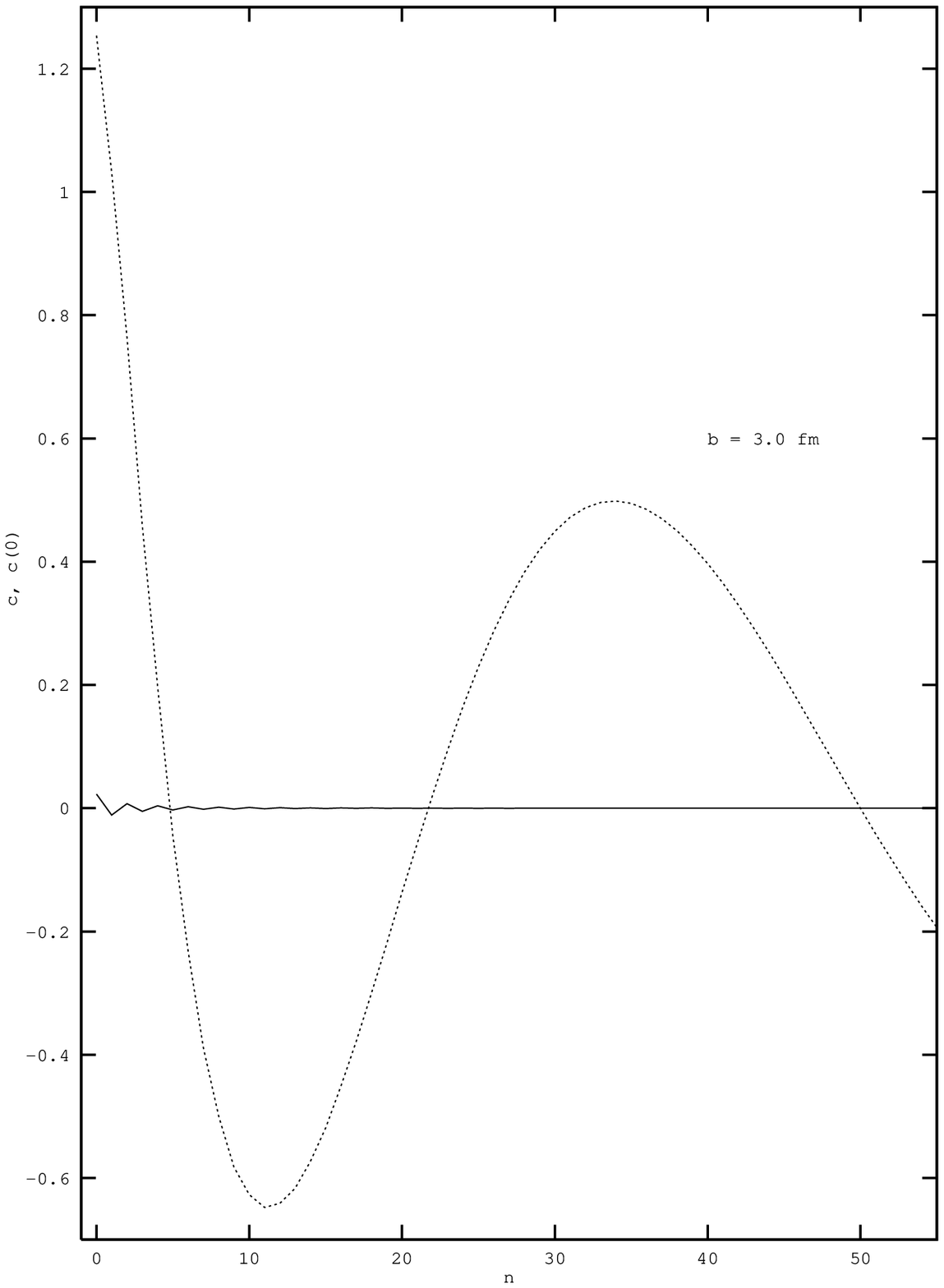}}
\caption{$c_n$ (dotted line) and $c_n^{(0)}$ (solid line) coefficients
obtained in a calculation involving 100 basis states, with $b = 3.0$ and an
energy of $1 \ MeV$.}
\label{fig:vect_comp_vs_n}
\end{figure}

To corroborate the fact that the reformulated AM equations (\ref
{eq:AlgEqWithS}) provide faster converging solutions, we show in Fig. \ref
{fig:vect_comp_vs_n} both the $c_n$ and $c_n^{(0)}$ coefficients obtained in
a calculation at $E=1 \ MeV$ and $b = 3.0$, with a total number of 100 basis
states. This picture confirms the fact that the true solution indeed
deviates only by a small amount from the asymptotic one.

\begin{figure}[tbp]
\centerline{\psfig{figure=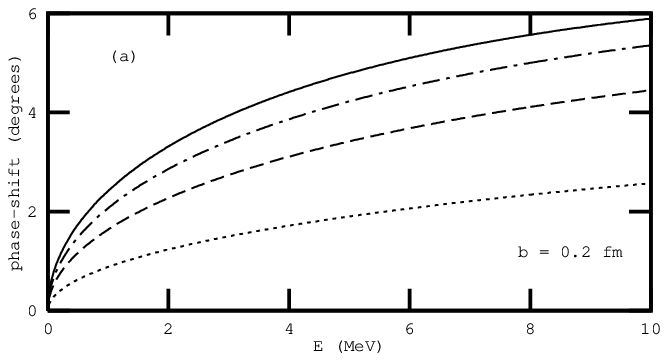} 
 \psfig{figure=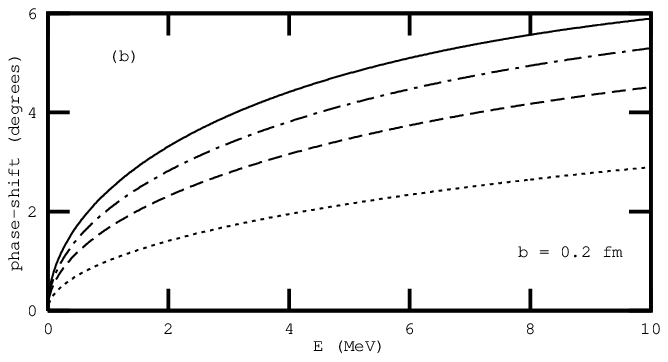}} 
\centerline{\psfig{figure=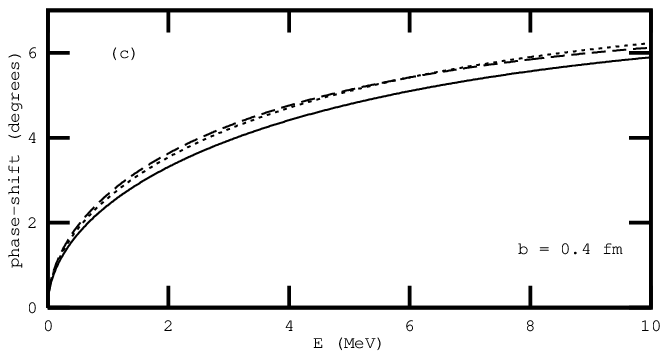} 
 \psfig{figure=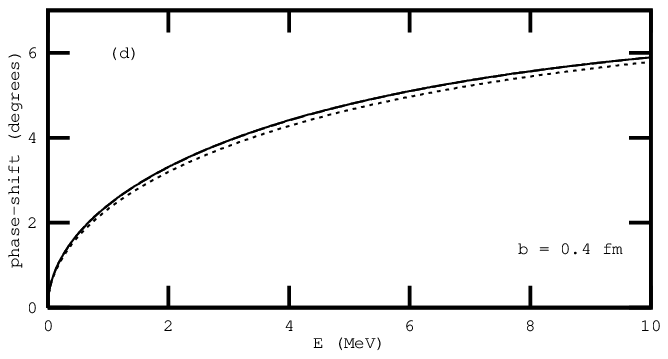}} 
\centerline{\psfig{figure=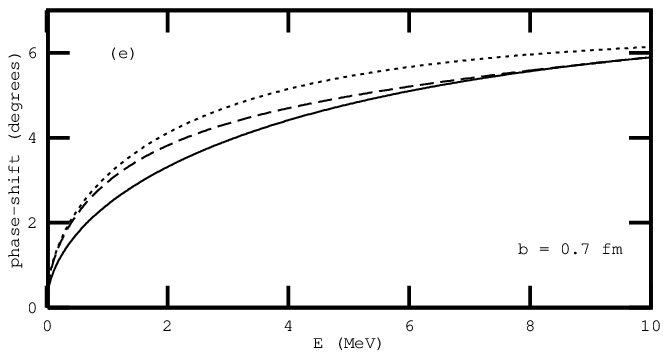} 
 \psfig{figure=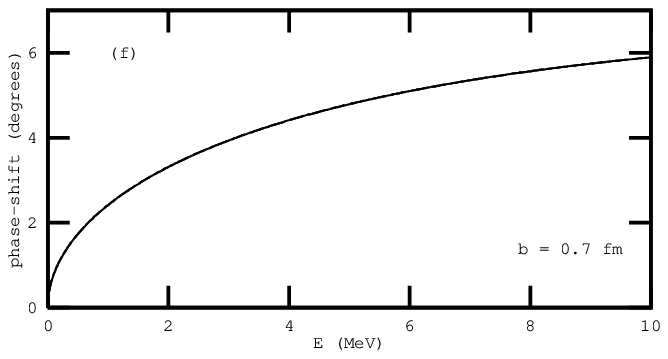}} 
\centerline{\psfig{figure=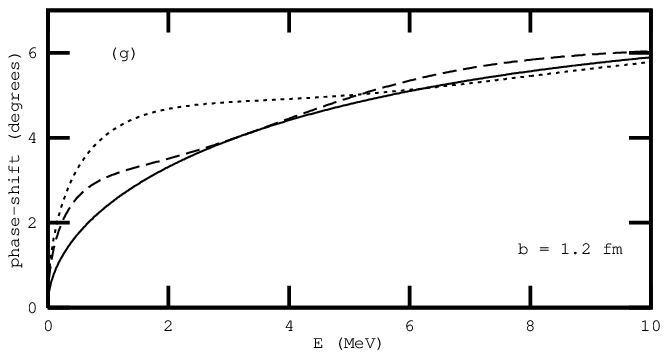} 
 \psfig{figure=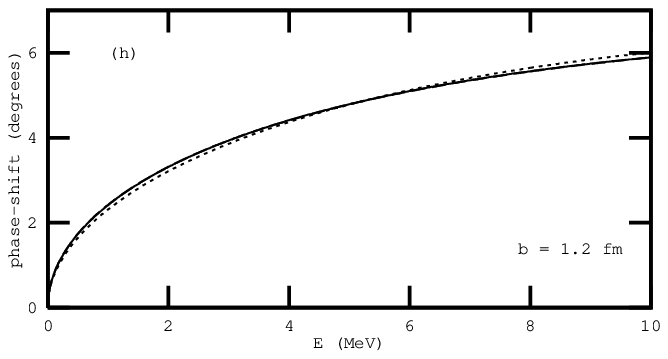}} 
\centerline{\psfig{figure=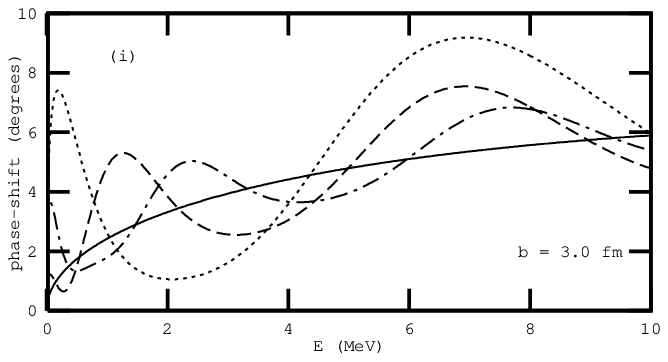} 
 \psfig{figure=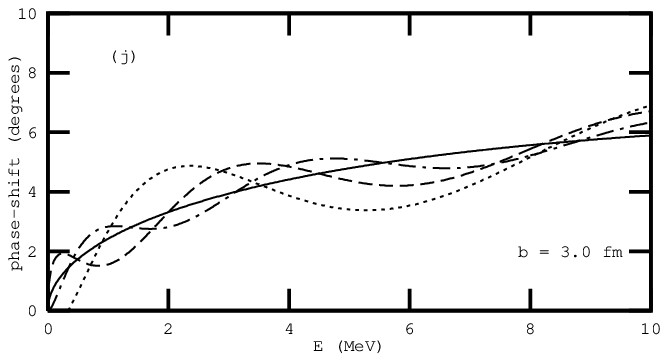}}
\caption{ Phase shift $\protect\delta$ (degrees) as a function of energy,
obtained with 5 (dotted line), 10 (dashed line) and 15 (dash-dotted line)
basis states. The solid line corresponds to the exact phase shift. Left side
figures: Phase shifts calculated with the original version of the dynamical
equations. Right side figures: Phase shifts calculated with the reformulated
version of the dynamical equations.}
\label{fig:phas_vs_e}
\end{figure}

To show the dependence of the results on energy, we show in Fig. \ref
{fig:phas_vs_e} the phase shift obtained with 5, 10 and 15 basis states
compared to the exact results, and this with the original (``simple'')
version (\ref{eq:AlgEqStraight}) and the reformulated version (\ref
{eq:AlgEqWithS}) of the AM equations. For $b$ values deviating reasonably
from the optimal value, the latter form of the equations is seen to be
highly superior to the original one. For very small and very large $b$
values, the convergence is still problematic.

\section{Stable Solutions for the AM Equations}

In the previous section we obtained interesting results concerning the
convergence behaviour of the solutions of the AM equations in terms of the
parameters defining the basis and the potential energy. It was shown for a
gaussian potential that the rate of convergence was reliably predictable.

The parameter $\gamma$ or, for fixed potential width $a$, the oscillator
radius $b$ determines the rate of convergence of the expansion $\Psi =
\sum_{n} c_{n} \mid n >$. It was indeed seen that an optimal value (or a
restricted range of optimal values) for $b$ leads to very fast convergence.
However, values deviating more or less strongly from the optimal $b$ lead to
slowly converging and numerically unprecise results.

As it is not always possible in a realistic calculation to choose the
optimal $b$, for physical as well as for numerical reasons, it is important
to develop strategies for stable results even in non-optimal $b$-regions.
This problem was already recognized in \cite{kn:alphacalc}, were a coupled
channels calculated for $^4He$ was performed with the AM, in which both
cluster and collective configurations were taken into account: the $b$ value
was fixed by physical arguments (essentially optimizing the cluster channel
results), and very non-optimal for the collective channels (in particular
the Monopole Mode).

The following subsections indicate how such problems can be solved by
considering novel strategies to solve the AM equations, without modifying
the original physical problem (e.g.\ by tampering with $b$).

\subsection{A General Analysis of the Asymptotics}

We attempt to evaluate the asymptotic behaviour (i.e.\ for $n \gg 1$) for
the expansion coefficients $c_n = <n \mid \Psi>$ and the matrix elements $<n
\mid \hat V \mid \Psi>$, using only very general information on the
(unknown) wave function $\Psi$. Based on this formulation we will then
introduce strategies to overcome the slow-convergence situations. All
assumptions made, and results obtained, will be checked against those
produced with a gaussian potential.

To derive a general view on the asymptotic behaviour of the Fourier
coefficients $c_n$, we use a Generator Coordinate (GC) representation for
the wave function $\Psi$: 
\begin{eqnarray}
\Psi(r) \, = \, \int_{0}^{\infty} \, d\beta \, g(\beta) \, r^L\, \exp\{ - {%
\beta}^2 r^2\}  \label{eq:GCform}
\end{eqnarray}
using a scaled gaussian $<r\mid\beta> = r^L\,\exp\{- {\beta}^2 r^2\}$ as a
kernel of the integral transformation. It is well known that this kernel
provides an alternative, continuous, basis to the traditional, discrete,
oscillator basis. A tacit assumption in (\ref{eq:GCform}) is that the
continuous spectrum wave functions are representable by such a
representation, (in particular, that a ``reasonable'' form for $g(\beta)$
exists, leading to an integrable result).

\subsubsection{Asymptotics of $c_n$}

The gaussian kernel of (\ref{eq:GCform}) is easily expanded in terms of the
oscillator basis used throughout this paper. The basis functions $\mid n>$
are explicited in a coordinate representation by (\ref{eq:OscFun}).

By substituting $\rho b$ for $r$ in the integral kernel $r^L\,\exp\{ - {\beta%
}^2 r^2$\}, one can recognizes the generating function for the oscillator
functions: 
\begin{eqnarray}
(1+\varepsilon)^{-L-3/2} \, {\rho}^L \exp \{-\frac{1}{2} \frac{1- \varepsilon%
}{1+\varepsilon} {\rho}^2 \} \, = \, \sum_{n=0}^{\infty} \frac{{\varepsilon}%
^{n}}{N_{n}} \mid n>
\end{eqnarray}
where 
\begin{eqnarray}
\beta^2 b^2 \, = \, \frac{1}{2} \frac{1-\varepsilon}{1+\varepsilon} \ \ \ %
\mbox{or} \ \ \ \varepsilon \, = \, \frac{1 - 2 {\beta}^2 b^2}{1 + 2 {\beta}%
^2 b^2}
\end{eqnarray}

The expansion coefficients $<n \mid \beta>$ of $\exp\{ - {\beta}^2 r^2\}$ in
the oscillator basis are then 
\begin{eqnarray}
<n\mid \beta> \, & = & \, (1+\varepsilon)^{L+3/2} \frac{{\varepsilon}^{n}}{%
N_{n}}\ b^L  \nonumber \\
& = & \, \frac{(1 - 2 {\beta}^{2} b^{2})^{n}}{(1 + 2 {\beta}^{2}
b^2)^{n+L+3/2}} \frac{2^{L+3/2}}{N_{n}}\ b^L  \label{eq:coeff}
\end{eqnarray}
from which one obtains the expansion coefficients $c_n = <n \mid \Psi \mid >$
of $\Psi$ in the oscillator basis as 
\begin{eqnarray}
<n\mid\Psi> \, = \, \int_{0}^{\infty} \, d\beta \, g(\beta) \, <n\mid\beta> 
\nonumber \\
= \int_{0}^{\infty} \, d\beta \, g(\beta) \frac{(1 - 2 {\beta}^{2} b^{2})^{n}%
}{(1 + 2 {\beta}^{2} b^2)^{n+L+3/2}} \frac{2^{L+3/2}}{N_{n}}
\label{eq:eqn125}
\end{eqnarray}
If, and when, $g(\beta)$ is concentrated in a small vicinity of $\beta
\approx {\beta}_0$, one can expect a highly convergent expansion for an
oscillator length $b \approx 1/\sqrt{2} \beta_0$.

To study the asymptotic behaviour of the expansion coefficients (\ref
{eq:eqn125}), we consider two limiting regions for $b$, i.e.\ ``small $b$''
and ``large $b$''. To this end, we rewrite (\ref{eq:eqn125}) as follows 
\begin{eqnarray}
<n\mid\Psi> \, = \, \int_{0}^{\beta_0} \, d\beta \, g(\beta) \, <n\mid\beta>
\, + \, \int_{\beta_0}^{\infty} \, d\beta \, g(\beta) \, <n\mid\beta>
\label{eq:eqn735}
\end{eqnarray}
where $\beta_0 = 1/(\sqrt{2}b)$.

\paragraph{Small $b$ values:}

for $0 \leq \beta < 1 / (\sqrt{2} b)$ and large values of $n$, one can use
the following approximate formula 
\begin{eqnarray}
\frac{(1 - 2 {\beta}^{2} b^{2})^{n}}{(1 + 2 {\beta}^{2} b^2)^{n+L+3/2}} \,
\approx \, \exp \{ - R_n^2 {\beta}^{2} b^{2} \}  \label{eq:543}
\end{eqnarray}
For large values of $n$ ($n \gg 1$) one also has 
\begin{eqnarray}
\frac{1}{N_{n}} \, = \, \sqrt{\frac{ \Gamma(n+L+3/2)}{2 \Gamma(n+1)}} \,
\approx \, R_n^{L+1/2}/ 2^{L+1}  \label{eq:544}
\end{eqnarray}
where again $R_n$ is the classical turning point (cfr. eqs (\ref
{eq:AlgCoefAsBound}) and (\ref{eq:AlgCoefAsCont})).

For very small values of $b$, one can consider only the first term in (\ref
{eq:eqn735}); using the approximations (\ref{eq:543}) and (\ref{eq:544}) one
then obtains the following asymptotic form of (\ref{eq:eqn125}) 
\begin{eqnarray}
<n\mid\Psi> \, \approx \, \sqrt{2} \sqrt{R_n} \int_{0}^{\infty} \, d\beta \,
g(\beta) \, (R_n b)^L \exp \left \{ - R_n^2 {\beta}^{2} b^{2} \right \}
\end{eqnarray}
which, considering (\ref{eq:GCform}), leads to the final approximation: 
\begin{eqnarray}
c_n \, = \, <n\mid\Psi> \, \approx \, \sqrt{2} \sqrt{R_n} \Psi (b \, R_n)
\end{eqnarray}
For small values of the oscillator length $b$, the expansion coefficients of
the wave function $\Psi(r)$ on an oscillator basis are thus proportional to
the wave function itself taken at the discrete argument values $b R_n$.

Such a relationship between the wave function in coordinate space and its
oscillator representation was already obtained long ago \cite{kn:Fil}, \cite
{kn:Heller1} for the regular and irregular asymptotic solutions. Later, this
correspondence has been used as a heuristic principle for solving e.g.\ the
Coulomb problem in an oscillator representation \cite{kn:Okhrim84}.

\paragraph{Large $b$ values:}

for $1 / (\sqrt{2} b) \leq \beta < \infty $ and large values of $n$, one can
use the following approximate formula 
\begin{eqnarray}
\frac{(1 - 2 {\beta}^{2} b^{2})^{n}}{(1 + 2 {\beta}^{2} b^2)^{n+L+3/2}} & =
& (-1)^n \frac{1}{(2{\beta}^2 b^2)^{L+3/2}} \frac{(1 - 1 / 2 {\beta}^{2}
b^{2})^{n}} {(1 + 1 / 2 {\beta}^{2} b^2)^{n+L+3/2}}  \nonumber \\
& \approx & (-1)^n \frac{1}{(2{\beta}^2 b^2)^{L+3/2}} \exp \{ - \frac{R_n^2}{%
4 {\beta}^{2} b^{2}} \}
\end{eqnarray}

The expansion coefficients can then be approximated by the second term in (%
\ref{eq:eqn735}), which for very large values of $b$ leads to the following
asymptotic form of (\ref{eq:eqn125}) 
\begin{eqnarray}
<n\mid\Psi> \, \approx \, (-1)^{n} \sqrt{2} R_n^{L+1/2}\ b^L
\int_{0}^{\infty} d\beta g(\beta) \frac{\exp \{ - R_n^2/4 {\beta}^{2} b^{2}
\}} {(2{\beta}^2 b^2)^{3/2}}
\end{eqnarray}
Using an integral transformation \cite[11.4.29]{kn:abra} the asymptotic form
of the $c_n$ for large $b$ becomes 
\begin{eqnarray}
c_n \, = \, <n\mid\Psi> & \, \approx \, & (-1)^{n} \frac{2}{\sqrt{\pi}} 
\sqrt{R_n} \int_{0}^{\infty} \, dr \, r^{2} \, j_{L} {\left(\frac{R_n}{b}%
r\right)} \, \Psi(r)  \label{eq:larg b 0} \\
& = & (-1)^{n} \frac{2}{\sqrt{\pi}} \sqrt{R_n} {\Phi}(\frac{R_n}{b})
\label{eq:larg b}
\end{eqnarray}
where $\Phi$ is the wave function in momentum representation.

For large values of the oscillator length $b$, the expansion coefficients of
the wave function $\Psi(r)$ on an oscillator basis are thus proportional to
the wave function in momentum space taken at discrete arguments which are
the momentum values $R_n / b$. As oscillator functions in coordinate and
momentum space only differ by a phase $(-1)^n$ and the argument $\rho$
(which in coordinate space equals $r/b$ and in momentum space equals $kb$),
both wave functions $\Psi(r)$ in coordinate representation and $\Phi(p)$ in
momentum representation have essentially the same expansion coefficients.

By integrating (\ref{eq:larg b 0}) by parts, assuming that $\Psi(r)$ is
smooth and without singular points, and taking the leading asymptotic term
for $c_n$ (i.e.\ for $R_n \gg 1$), one obtains 
\begin{eqnarray}
<n \mid \Psi> \, \approx \, (-1)^{n} \frac{1}{R_{n}^{3/2}} \Psi(0)
\label{eq:eq333}
\end{eqnarray}

The same result can be obtained by considering that the wave functions $%
\Phi(p)$ of discrete and continuous spectrum states in momemtum space
decrease at least as $1/ p^2$. Substitution of this limiting factor in (\ref
{eq:larg b}) again leads to (\ref{eq:eq333})

Summing up one notes that the asymptotics of $c_n$ are related to the wave
function behaviour in coordinate space: for small values of $b$ the
asymptotics are proportional to the wave function for large $r$, and for
large values of $b$ the asymptotics are proportional to the wave function in
the vicinity of the origin.

\subsubsection{Asymptotics of $<n \mid \hat V \mid \Psi>$}

Asymptotic expressions for the matrix elements $< n \mid \hat{V} \mid \Psi >$
can be obtained in an analogous way as for $c_n = < n \mid \Psi >$ by using
the GC representation of the solution (\ref{eq:GCform}). We omit the details
of the calculation, and only present the final results.

\paragraph{Small $b$ values:}

for small $b$ values (or better, small $\gamma$ values), the matrix elements
factorize as follows: 
\begin{eqnarray}
< n \mid \hat{V} \mid \Psi > & \approx & \hat{V} (b R_n) \ \ < n \mid \Psi >
\nonumber \\
& \approx & \hat{V} (b R_n) \sqrt{2} \sqrt{R_n} \Psi (b R_n)
\label{eq:Vasysmall}
\end{eqnarray}
One notices that the behaviour of the matrix elements coincides with the one
of the $\Psi (r)$, up to a trivial factor which is potential dependent.

These results obtained for a general form of a short-range potential, are
confirmed by the asymptotic form for $V_{n}^{(+)}$, evaluated earlier (\ref
{eq:V_plus_asym_small_b}) for a gaussian potential.

\paragraph{Large $b$ values:}

for large values of $b$ (or $\gamma$), the asymptotic form of the matrix
elements $< n \mid \hat{V} \mid \Psi >$ reduces to the integral 
\begin{eqnarray}
< n \mid \hat{V} \mid \Psi > \approx (-1)^n \frac{2}{\sqrt{\pi}} \sqrt{R_n}
\int_{0}^{\infty} dr r^2 j_{L}(\frac{R_n}{b}r) \hat{V} (r) \Psi(r)
\label{eq:eq698}
\end{eqnarray}
or, in other words, to the convolution of the wave function with the
potential in momentum representation 
\begin{eqnarray}
< n \mid \hat{V} \mid \Psi > \approx (-1)^n \frac{2}{\sqrt{\pi}}\sqrt{R_n}
\int_{0}^{\infty} d\tilde{k} \, \tilde{k}^2 \, \hat{v} (k_n - \tilde{k}) \, {%
\Phi}(\tilde{k})
\end{eqnarray}
where $\hat{v}(k)$ and ${\Phi}_{L}(k)$ are the Fourier transforms of the
potential $\hat{V}(r)$ and the wave function ${\Psi} (r)$ respectively.

These results for a general potential can again be checked by considering
the corresponding form for $V_{n}^{(+)}$ obtained earlier with a gaussian
potential. Indeed, by calculating the integral 
\begin{eqnarray}
V_{0} \int_{0}^{\infty} dr r^2 j_{L} (\frac{R_n}{b}r) \exp \left \{ -\gamma
r^2 \right \} j_{L} (kr)
\end{eqnarray}
where the unknown solution $\Psi(r)$ is substituted by $\Psi^{(+)} (k, r) = 
\sqrt{2/\pi} j_L(kr)$ (the ``free particle'' solution), one obtains (\ref
{eq:Spluslargeb}), which is valid for both large $b$ and $n$.

The matrix elements $< n \mid \hat{V} \mid \Psi >$ can be evaluated in the
asymptotic region by making the same assumptions concerning the behaviour of 
$\Psi (r)$ as in the previous section to obtain the asymptotic form of $c_n
= <n \mid \Psi >$. Expanding $\Psi (r)$ in a power series of $r$, keeping
the first term $\Psi (0) \, r^L$ and integrating term by term in (\ref
{eq:eq698}) leads to the following asymptotic form: 
\begin{equation}
< n \mid \hat V \mid \Psi> \approx (-1)^n \frac{2}{\sqrt{\pi}} \sqrt{R_n}
\hat v (\frac{R_n}{b}) \Psi (0)  \label{eq:Vasymlargb}
\end{equation}
where the Fourier transform $\hat v (R_n / b)$ of the potential determines
how fast the matrix elements decrease in terms of $n$.

\subsection{New Strategies for Solving the AM Equations}

Based on the results obtained above, we can now suggest new strategies for
solving the dynamical equations of the AM, in those cases where the
oscillator length $b$, or more exactly the ratio $\gamma = (b / a)^2$, is
relatively large or small compared to the optimal value.

\paragraph{Small $b$ values:}

we start from the Schr\"{o}dinger equation in (\ref{eq:AlgEqGen}) 
\begin{eqnarray}
\sum_{m} < n \mid \hat{T} - E \mid m > c_m + <n \mid \hat{V} \mid \Psi > = 0
\\
\sum_{\mu} < \nu \mid \hat{T} - E \mid \mu > c_\mu + <\nu \mid \hat{V} \mid
\Psi > = 0
\end{eqnarray}
where the index $n$ is connected with the internal region and $\nu$ with the
asymptotic region, and $N$ delimits the internal region. In the asymptotic
region, we can use the asymptotic form (\ref{eq:Vasysmall}) for the matrix
elements 
\begin{equation}
<\nu \mid \hat{V} \mid \Psi > = \hat{V} (b R_{\nu}) \, c_{\nu}
\end{equation}
which yields 
\begin{eqnarray}
\sum_{m} < n \mid \hat{T} - E \mid m > c_m + <n \mid \hat{V} \mid \Psi > = 0
\\
\sum_{\mu} < \nu \mid \hat{T} + \hat{V}(b R_{\nu}) - E \mid \mu > c_\mu = 0
\label{eq:eq201}
\end{eqnarray}

In the original version of the AM the asymptotic coefficients $c_{\nu}$ were
given by (\ref{eq:casytrad}), and obtained from a simple three term
recurrence relation, due to the very selective coupling induced by the
kinetic energy operator between oscillator states. Instead of the original
form, and based on (\ref{eq:eq201}), we propose a modified three term
recurrence form, including the asymptotic behaviour of the potential matrix
elements in the diagonal term 
\begin{eqnarray}
\sum_\mu < \nu \mid \hat{T} + \hat{V}(b R_\nu) - E \mid \mu > c_\mu = 0
\end{eqnarray}
or 
\begin{eqnarray}
& & [< \nu \mid \hat{T} \mid \nu > \, + \, \hat{V}(b R_{\nu}) \, - \, E]
c_{\nu}  \nonumber \\
& & + \, < \nu \mid \hat{T} \mid \nu-1> c_{\nu-1} \, + \, < \nu \mid \hat{T}%
\mid \nu+1> c_{\nu+1} = 0  \label{eq:eqn120}
\end{eqnarray}

In the asymptotic region $c_{\nu} = c_{\nu}^{(+)} + \tan \delta c_{\nu}^{(-)}
$, hence the system (\ref{eq:eqn120}) should be solved independently for
regular ($c_{\nu}^{(+)}$) and irregular ($c_{\nu}^{(-)}$) coefficients, with
"boundary" conditions at starting remote points $\nu = N_a, N_a + 1$ 
\begin{eqnarray}
c_{{\nu}_0}^{(+)} = \sqrt{2 R_{\nu}} j_{L}(k R_{\nu}) \\
c_{{\nu}_0}^{(-)} = \sqrt{2 R_{\nu}} n_{L}(k R_{\nu})
\end{eqnarray}
These modified asymptotic coefficients $c_{\nu}^{(+)}$ and $c_{\nu}^{(-)}$
should be considered when calculating the correspondingly modified $V_n^{(+)}
$ and $V_n^{(-)}$ in the dynamical equations (\ref{eq:AlgEqWithS}). The
modified equations should then be solved for the phase shift $\delta$ and
the coefficients of the internal region $c_n^{(0)}$.

\paragraph{Large $b$ values:}

for large values of $b$, we consider the dynamical equations in the form (%
\ref{eq:AlgEqNew}), containing the $V_n^{(+)}$ and $V_n^{(-)}$ terms

\begin{eqnarray}
\sum_{m=0} <n \mid \hat{H} - E \mid m> c_m^{(0)} + \, V_n^{(-)} \,
\tan(\delta) \, = \, - \, V_n^{(+)}  \label{eq:eqn121}
\end{eqnarray}
for the internal region ($n \le N$), and 
\begin{equation}
\sum_{\mu} < \nu \mid \hat T - E \mid \mu > c_\mu + < \nu \mid \hat V \mid
\Psi > = 0  \label{eq:eq743}
\end{equation}
for $\nu > N$.

Substituting $c_{\mu}^{(0)} + c_{\mu}^{(+)} + \tan (\delta) c_{\mu}^{(-)}$
for $c_\mu$ in (\ref{eq:eq743}), an assuming that $c_{\mu}^{(+)}$ and $%
c_{\mu}^{(-)}$ satisfy the three-term recurrence relation (\ref{eq:AsCoefDef}%
), one obtains 
\begin{equation}
\sum_{\mu} < \nu \mid \hat T - E \mid \mu > c_\mu^{(0)} + < \nu \mid \hat V
\mid \Psi > = 0
\end{equation}
Taking the asymptotic form (\ref{eq:Vasymlargb}) into account, one obtains
the following equation for $c_\mu^{(a)}$ 
\begin{equation}
\sum_{\mu} < \nu \mid \hat T - E \mid \mu > c_\mu^{(a)} + (-1)^n \sqrt{R_{\nu}}
v(\frac{R_{\nu}}{b}) = 0  \label{eq:eq945}
\end{equation}
where $c_\nu^{(0)}$ was substituted by 
\begin{equation}
c_\nu^{(0)} = c_\nu^{(r)} + W \, c_\nu^{(a)}  \label{eq:eq007}
\end{equation}
introducing the "asymptotic" coefficients $c_\nu^{(a)}$ and the "residual"
coefficients $c_\nu^{(r)}$. All constants in the asymptotic form (\ref
{eq:Vasymlargb}) were omitted and replaced by the global factor W to be
determined. Equation (\ref{eq:eq945}) should now be solved subject to the
boundary condition 
\begin{equation}
c_\nu^{(a)} = 0
\end{equation}
for $\nu = N_a, N_a + 1$. Having obtained $c_\nu^{(a)}$, the equations for
the internal region should be modified to take the substitution (\ref
{eq:eq007}) into account 
\begin{eqnarray}
\sum_{m=0} <n \mid \hat{H} - E \mid m> c_m^{(r)} \, + \, \left [ V_n^{(a)} -
E c_n^{(a)} \right ] \, W \, + \, V_n^{(-)} \, \tan(\delta) \, = \, - \,
V_n^{(+)}
\end{eqnarray}
were $V_{n}^{(a)}$ now stands for the sum 
\begin{equation}
V_n^{(a)} \, = \, \sum_{m=0} \, <n \mid \hat{H} \mid m> c_m^{(a)}
\end{equation}

The meaning of the asymptotic coefficients $c_{n}^{(a)}$ can be most easily
understood by considering the major part of the asymptotic behaviour already
in an intermediate $n$ region. Indeed, for large $b$ values, there is a very
slow convergence of the results. This implies that the internal region
extends towards very large $n$, with corresponding $c_n^{(0)}$ solutions
probably very close (i.e.\ to first order) to the $c_{n}^{(a)}$.

\paragraph{Numerical application:}

To demonstrate how the new (``small $b$'' and ``large $b$'') strategies
accelerate the convergence, we again consider a gaussian potential in the
two limiting cases where (i) the oscillator radius is a factor of 5 less
than the potential range, and (ii) the oscillator radius is a factor of 5
larger than the radius of the potential. 
\begin{figure}[tbp]
\centerline{\psfig{figure=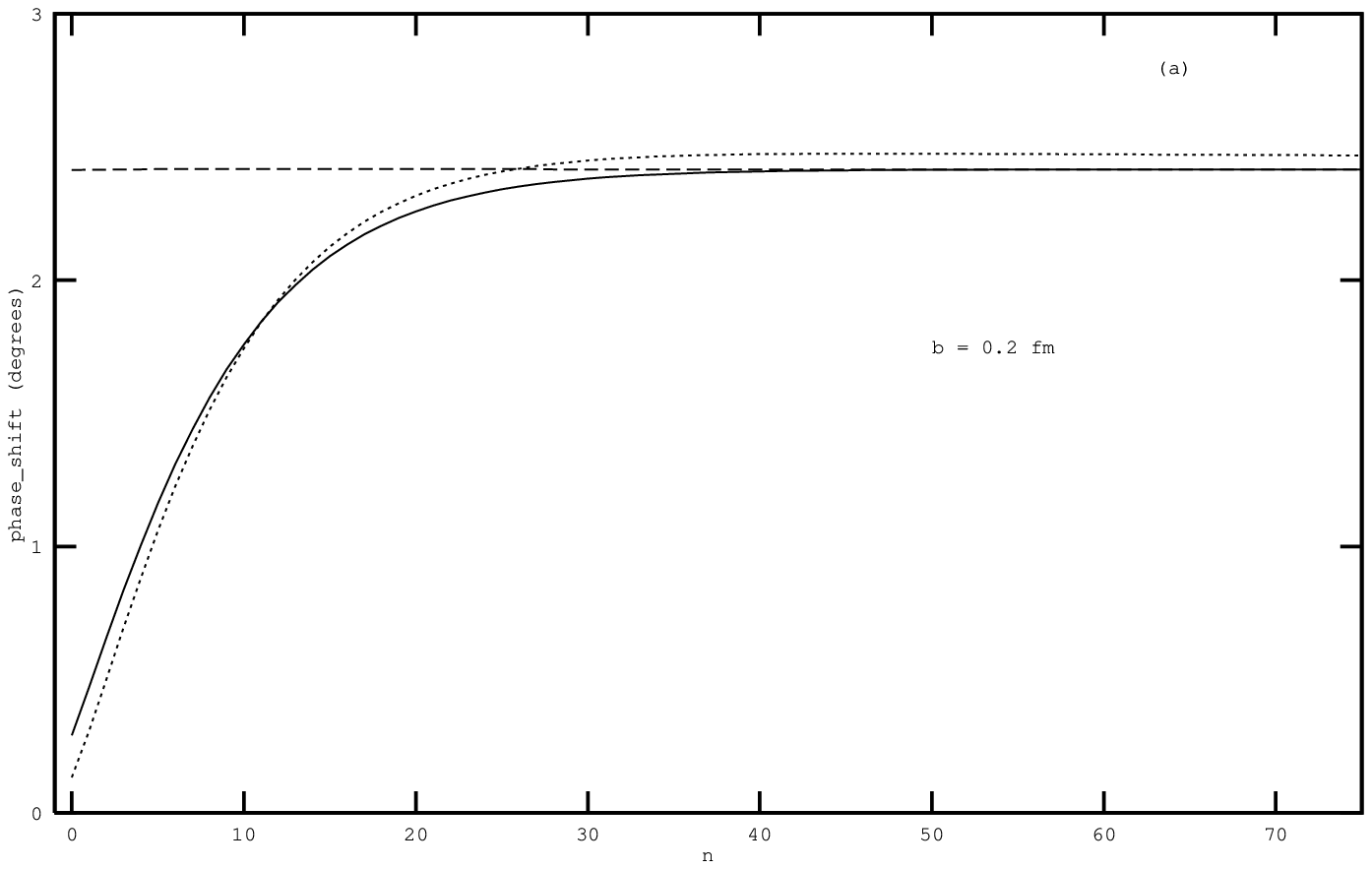}} \centerline{\psfig{figure=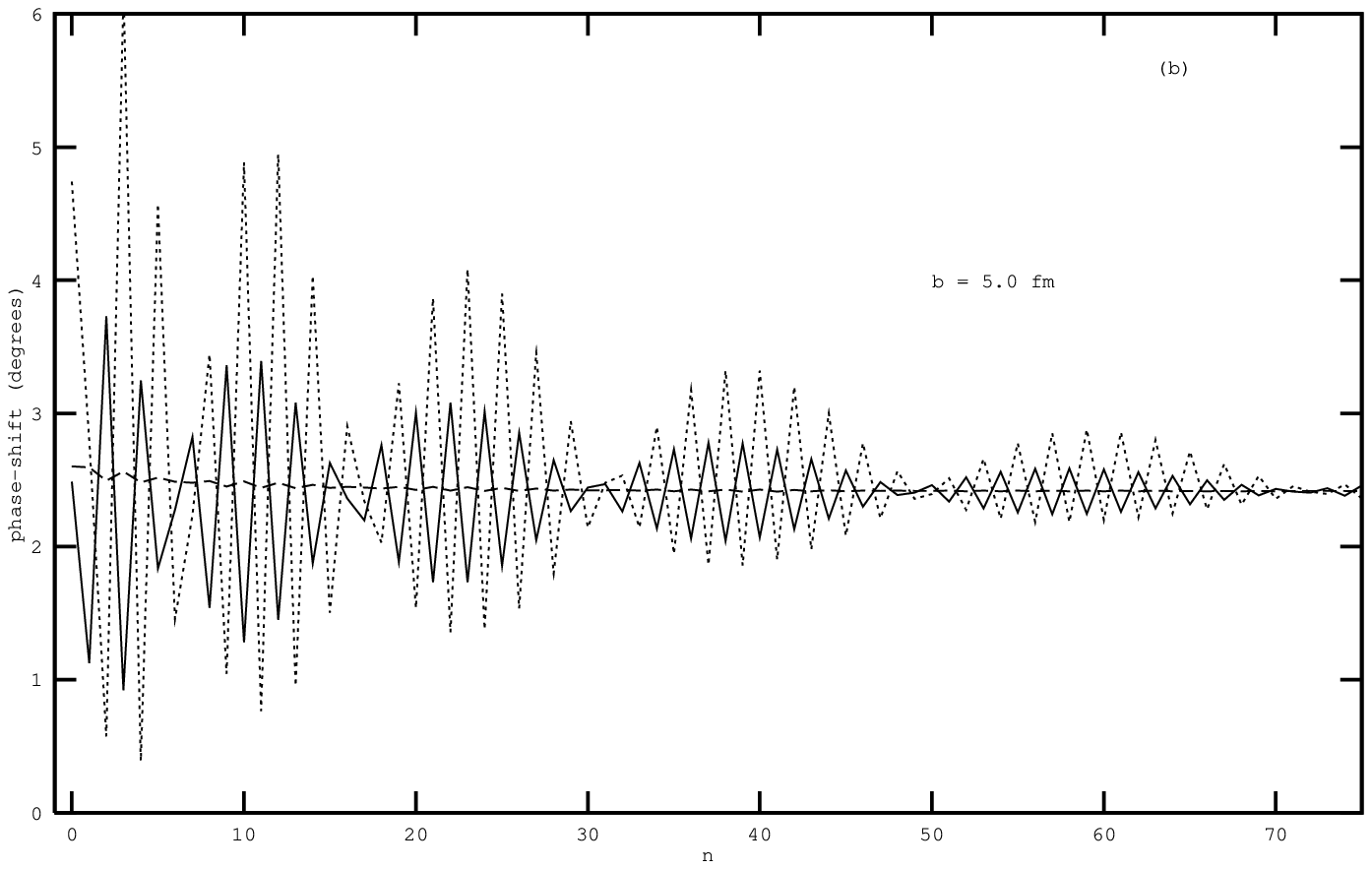}}
\caption{Phase shifts at an energy of $1 \ MeV$ as a function of the number
of basis states involved in the calculation. Dotted lines represent
solutions of the original (``simple'') formulation, solid lines represent
solutions of the reformulated version and dashed lines represent solutions
of the ``new strategies''.}
\label{fig:phas_vs_n_method}
\end{figure}

Fig. \ref{fig:phas_vs_n_method} compares the phase shifts, obtained at an
energy of $1 \ MeV$, obtained in the original, reformulated and asymptotic
approaches.

\begin{figure}[tbp]
\centerline{\psfig{figure=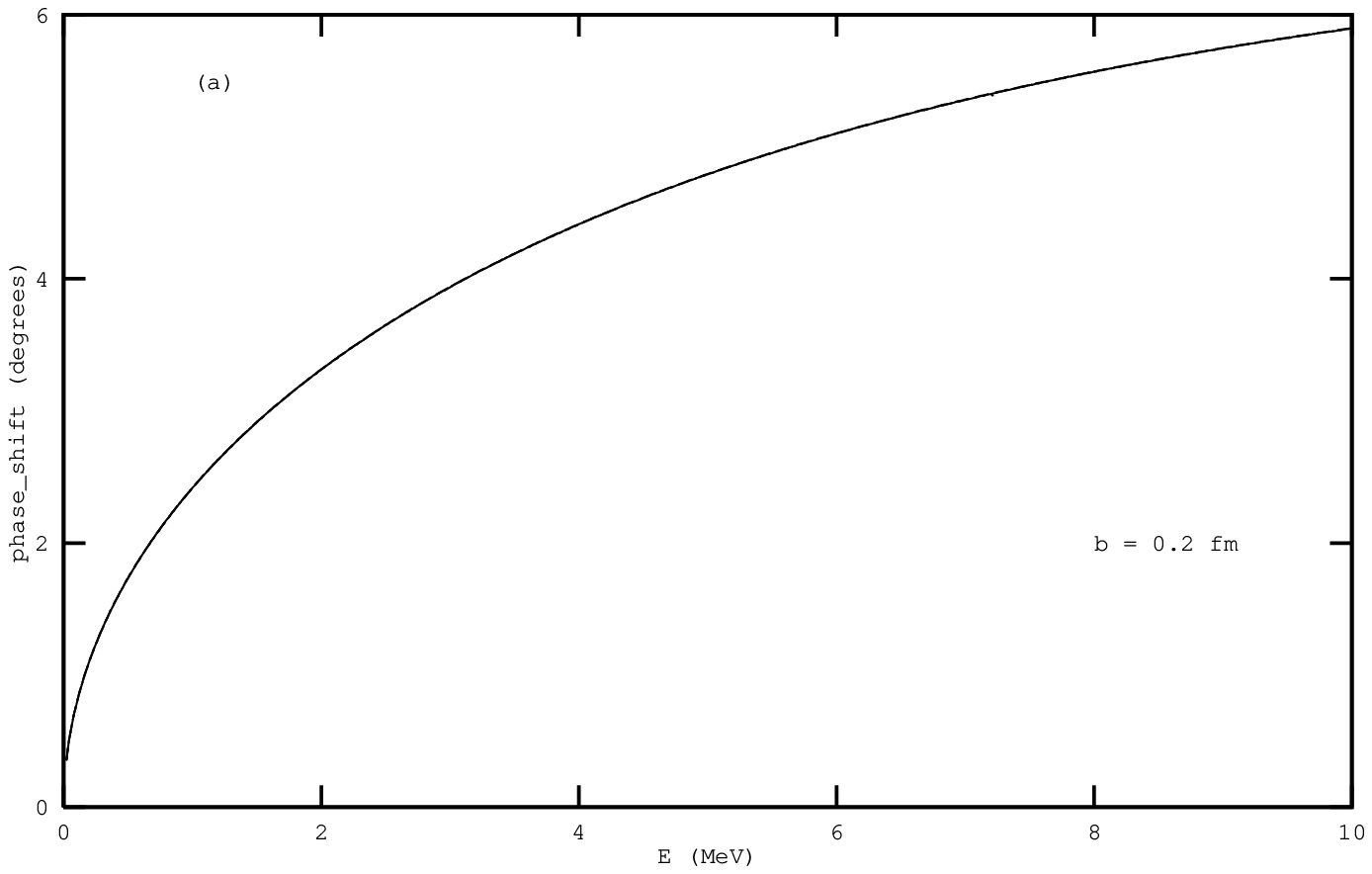}} \centerline{\psfig{figure=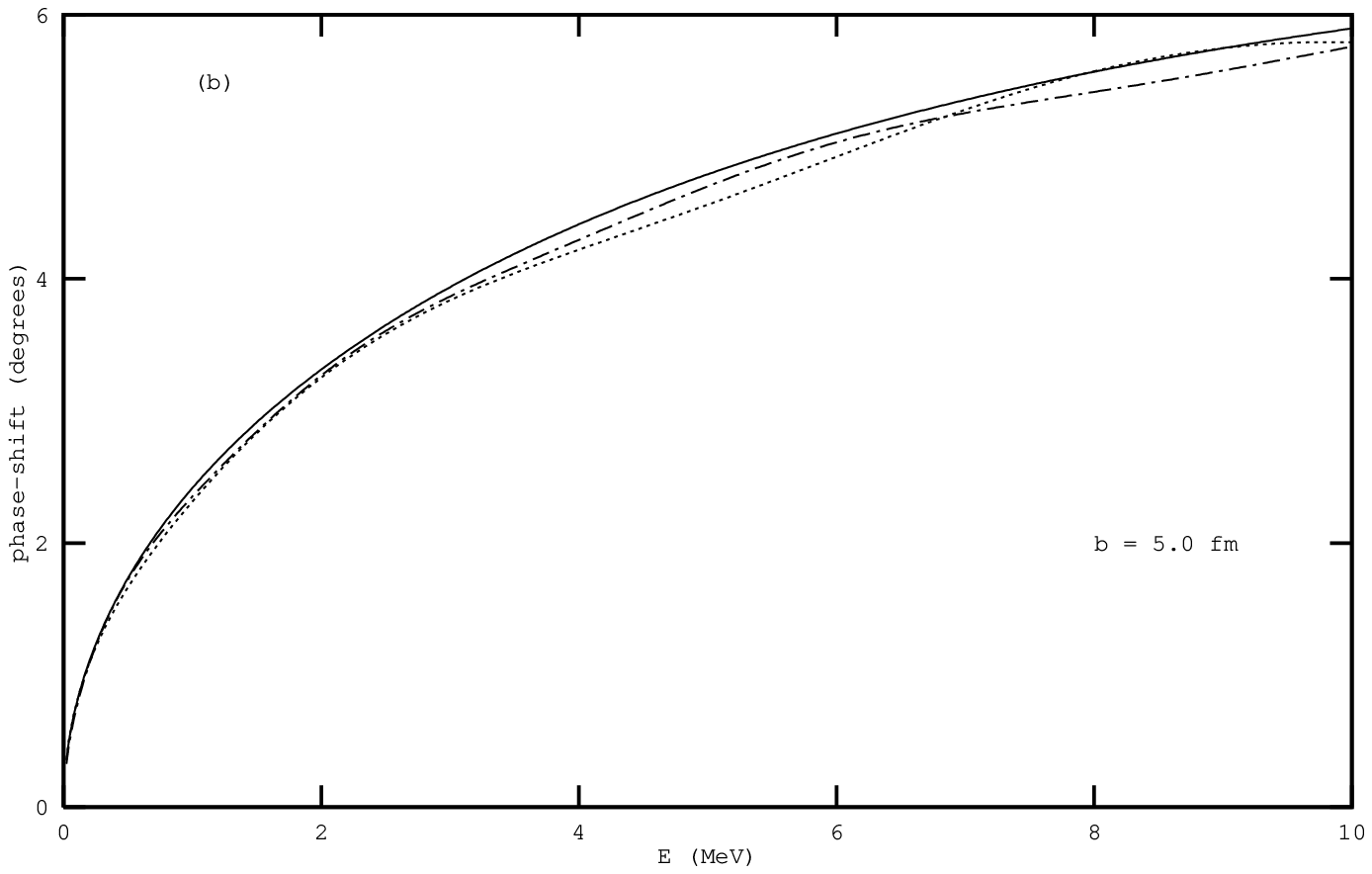}}
\caption{Phase shifts in the energy range from 0 to $10 \ MeV$ for a ratio $%
b / a = 0.2$, obtained in the ``small $b$ strategy'' (top), and $b / a = 5$,
obtained in the ``large $b$ strategy'' (bottom). Full lines show the exact
solution, dotted (resp. dashed) solutions are obtained with 5 (resp. 10)
basis states.}
\label{fig:phas_vs_e_method}
\end{figure}

In Fig. \ref{fig:phas_vs_e_method} we display the exact phase shifts and
those obtained with the new strategies using 5, respectively 10, basis
states. One notices that for the ``small $b$'' case, 5 basis functions yield
almost exact results (within a precision less than $0.01 \%$); by even
considering only one basis function, the phase shift is obtained within a
precision of $1 \%$! In the original form the latter precision could only be
reached by using more than 50 basis functions. 
For the ``large $b$ strategy'' ($b/a=5$), the convergence speed is
remarkably increased, although less spectacular than in the ``small $b$''
case. The number of states needed for an identical result in the original
formulation (\ref{eq:AlgEqWithS}) is about 5 times higher.

The two strategies suggested are thus seen to significantly improve the
convergence of the results and to reduce the computational efforts to obtain
a desired precision.

By modifying the form of the asymptotic AM equations through inclusion of
dynamical features as proposed and realized above for both small and large
values of $b$, one is naturally led to the introduction of an ``intermediate
region''. This region distinguishes itself (i) from the ``internal region''
where the solutions are governed by the potential, and (ii) from the
``asymptotic region'' where the kinetic energy dominates the equations. In
the intermediate region, the solutions are easily (i.e.\ in a numerically
simple way) obtained. There are then actually two variational parameters to
consider, $N$ marking the border between the internal and intermediate
region, and $N_a$ marking the transition from the intermediate to the true
asymptotic region. The larger or smaller is $\gamma$, the larger is the size
(i.e.\ the difference $N_a - N$) of the intermediate region. The strategies
proposed above for small and large values of $\gamma$ have shown that, by
considering a sufficiently large intermediate region (e.g.\ $N_a = 500$),
significantly reduces the range of the internal region, and dramatically
decreases the computational effort to obtain converged results.

The intermediate region is charactarised by a noticeable but non-dominant
presence of dynamical (i.e.\ potential) effects. These effects can be well
approximated by asymptotic forms of both the expansion coefficients $c_n$
and potential matrix elements $< n \mid \hat{V} \mid \Psi>$. For small $%
\gamma$ (long-range correlations), this leads to a redefinition of the
equations of $c_n^{(+)}$ and $c_n^{(-)}$ by incorporation a potential term
in the three term recurrence relation governing the true asymptotic
behaviour. For large $\gamma$ (short-range correlations) one approximates
part of the $c_n^{(0)}$ solution by its asymptotic behaviour $c_n^{(a)}$.
The latter are solutions of an inhomogeneous set of linear equations
containing the Fourier transform of the potential, and as such accumulate
the potential effects in the intermediate region.

In the region of $\gamma$ near to the optimal value, where fast converging
solutions are obtained, there is no need for an intermediate region, and
thus for an approximation of the dynamical equations; the values of $N$ and $%
N_a$ coincide in this case.

\section{Conclusions}

By introducing the ``Dynamical Coefficients'' $V_n^{(+)}$ and $V_n^{(-)}$ we
were able to suggest a new form of the dynamical equations of the Algebraic
Version of the Resonating Group Method (AM). The dynamical coefficients
allow one to determine oscillator basis parameters (oscillator length and
number of basis functions) for obtaining otimally converging and stable
solutions of these equations. It was shown that for an optimal parameter set
5 to 10 functions were sufficient to obtain results with a precision of more
than $99.9 \%$.

The new form of equations have also been shown to be a proper starting point
for obtaining new solution strategies in parameter regions were slow
convergence would occur. Indeed, in parameter regions were the oscillator
length is much larger, respectively much smaller, than the optimal one
asymptotic approximations could be formulated by analysing the AM equations.
The latter can be easily implemented numerically, and have shown to
dramatically improve the convergence of the solutions.

It was shown also that, if the oscillator radius is much smaller than the
width of the potential, the expansion coefficients $c_n$ of the wave
function on the oscillator basis coincide with the wave function in
coordinate space up to a simple factor. In the other limiting case, when the
oscillator radius is much larger than the width of the potential, the
expansion coefficients $c_n$ were found to be proportional to the wave
function in momentum space.

%\input{acknowledg}
%\acknowledgments

\section{Acknowledgments}

The authors are gratefull to Profs. ~G.~F.~Fillipov and ~P.~Van Leuven for
valuable discussions. They would also like to thank INTAS for financial
support for part of this work under the grant contract INTAS 93/755. One of
us (V.\ S.\ V.) gratefully acknowledges the kind hospitality of the members
of the research group ``Computational Qauntum Physics'' of the Department of
``Mathematics and Computer Sciences'', University of Antwerp, RUCA, Belgium,
and the ``Departement voor Wetenschappelijke, Technische en Culturele
aangelegenheden, Deelregering Vlaanderen'' for a scholarship.

\appendix
\section{appendix}

%\end{verbatim}

In this appendix the explicit form, and governing recurrence relations, of
the dynamical coefficients $V_{n}^{(+)}$ and $V_{n}^{(-)}$ are calculated
for a gaussian potential.

The following integrals should be evaluated: 
\begin{eqnarray}
V_{n}^{(+)} & = & V_{0} \, \int_{0}^{\infty} dr r^2 {\Phi}_{n} (r) \exp
\left (- r^2 / a^2 \right ) \, {\Psi}^{(+)} (kr)  \label{eq:app001} \\
V_{n}^{(-)} & = & V_{0} \, \int_{0}^{\infty} dr r^2 {\Phi}_{n} (r) \exp
\left (- r^2 / a^2 \right) \, {\Psi}^{(-)} (kr)  \label{eq:app002}
\end{eqnarray}
where ${\Psi}^{(+)} = \sqrt{\frac{2}{\pi}} j_L (kr)$ and ${\Psi}^{(-)}$ is a
solution of the inhomogeneous differential equation: 
\begin{eqnarray}
(\hat{T} - E) {\Psi}^{(-)} = {\beta}_0 {\Phi}_0 (r)
\end{eqnarray}
($\beta_0$ is defined in eq. (\ref{eq:beta}) ), which can be represented in
the integral form 
\begin{eqnarray}
{\Psi}^{(-)} & = & {\beta}_0 \int d\tilde{r} {\tilde{r}}^2 G(r, \tilde{r}) {%
\Phi}_0 (\tilde{r})  \nonumber \\
& = & {\beta}_0 \frac{2}{\pi}\int d\tilde{k} \int d\tilde{r} {\tilde{r}}^2 
\frac{j_L(\tilde{k}r) j_L(\tilde{k}\tilde{r})} {k^2-{\tilde{k}}^2} {\Phi}_0 (%
\tilde{r})  \nonumber \\
& = & {\beta}_0 \sqrt{\frac{2}{\pi}} \int d\tilde{k} \frac{j_L(\tilde{k}r)
c_{0}^{(+)}(\tilde{k})} {k^2-{\tilde{k}}^2}  \label{eq:PsiMin}
\end{eqnarray}

Integral (\ref{eq:app001}) is easily obtained by using formula 7.421.4 from 
\cite{kn:grad} yielding 
\begin{eqnarray}
V_{n}^{(+)} & = & (-1)^{n} V_{0} \frac{(1-2\gamma)^{n}} {(1+2
\gamma)^{n+L+3/2}} q^{L} \exp(-\frac{1}{2}k^{2}\frac{1}{1+2 \gamma}) N_{nL}
L_{n}^{L+1/2}(\frac{k^{2}}{1- 4 \gamma^{2}})  \nonumber \\
& = & V_{0} \frac{(1-2 \gamma)^{n+L/2}}{(1+2 \gamma)^{n+(L+3)/2}} \exp(\frac{%
1}{2} k^{2} \frac{\gamma}{1+2 \gamma}) c_{n}^{(+)} (k/\sqrt{1- 4 \gamma^2})
\label{eq:S+}
\end{eqnarray}
If $\gamma = \frac{1}{2}$, then 
\begin{eqnarray}
V_{n}^{(+)} = V_{0} N_{nL} \frac{1}{4} \frac{1}{n!} (\frac{k}{2})^{2n+L+1/2}
\exp(-\frac{1}{4} k^{2})
\end{eqnarray}

The second integral (\ref{eq:app002}), reduced to the form 
\begin{eqnarray}
V_{n}^{(-)} = {\beta}_0 \int d\tilde{k} \frac{V_{n}^{(+)}(\tilde{k})
c_{0}^{(+)} (\tilde{k})}{k^2-\tilde{k}^2}
\end{eqnarray}
can be expressed through the $c_{n}^{(-)}$ coefficients 
\begin{eqnarray}
V_{n}^{(-)} \, = \, V_{0} \frac{\exp(\frac{1}{2} q^{2} \frac{\gamma}{2
\gamma+1})} {(1+2 \gamma)^{n+(L+3)/2} (1+ \gamma)^{n+L/2}}  \nonumber \\
\sum_{m=0}^{n} \frac{ \Gamma(n+L+3/2)}{m! \Gamma(n-m+L+3/2)} \frac{N_{nL}}{%
N_{n-m,L}} [\gamma(1+2\gamma)]^{m} c_{n-m}^{(-)}(q)
\end{eqnarray}
where 
\begin{eqnarray}
q = k \sqrt{\frac{1 + \gamma}{1+2 \gamma}}
\end{eqnarray}
New variables $k^{\prime}$ and $\tilde{k}^{\prime}$ 
\begin{eqnarray}
k^{\prime}= \sqrt{\frac{1+\gamma}{1+2 \gamma}} \,k, \tilde{k}^{\prime}= 
\sqrt{\frac{1+\gamma}{1+2 \gamma}}\, \tilde{k},
\end{eqnarray}
and the following relation for Laguerre polynomials (see \cite{kn:Erdelyi},
vol. 2, formula 10.12(40)) 
\begin{eqnarray}
L_{n}^{\alpha}(\lambda x) = \sum_{m=0}^{n} \frac{\Gamma(n+\alpha+1)}{m!
\Gamma(n-m+\alpha+1)} \lambda^{n-m} (1-\lambda)^{m} L_{n-m}^{\alpha}(x)
\end{eqnarray}
were introduced in (\ref{eq:PsiMin}) to obtain this result.

To obtain the recurrence relations for $V^{(+)}_{n}$ and $V^{(-)}_{n}$, we
start from the equations for the functions $\Psi^{(+)}$ and $\Psi^{(- )}$ 
\begin{eqnarray}
(\hat{T} - E) \Psi^{(+)} & = & 0 , \\
(\hat{T} - E) \Psi^{(-)} & = & \beta_0 \Phi_{0} (r)
\end{eqnarray}
Multiplying both sides of these equations by $\Phi_{n}(r) V(r)$ and
integrating, one obtains 
\begin{eqnarray}
<n | \hat{V} \hat{T} | \Psi^{(+)} > - E V^{(+)}_{n} & = & 0, \\
<n | \hat{V} \hat{T} | \Psi^{(-)} > - E V^{(-)}_{n} & = & \beta_0 <n \mid 
\hat{V} \mid 0>
\end{eqnarray}
On notices the similarity of the left hand side of both equations, leading
to identical recurrence relations for both type of coefficients. The matrix
element $<n \mid \hat{V} \mid 0> $ for the gaussian potential is 
\begin{eqnarray}
<n \mid \hat{V} \mid 0> = (-1)^n V_{0} (1-z)^n z^{L+3/2} \sqrt{\frac{%
\Gamma(n+L+3/2)}{n! \Gamma(L+3/2)}}
\end{eqnarray}
where 
\begin{eqnarray}
z = (\gamma + 1)^{-1}
\end{eqnarray}
Using the commutation relations of the operators $\hat{T}$ and $\hat{V}$, or
the explicit form of $V_{n}^{(+)}$, one finally obtains 
\begin{eqnarray}
(1+ 2\gamma)^{2} T_{n,n+1} V^{(+)}_{n+1} + [(1 - 4\gamma^{2}) T_{n,n} - E ]
V^{(+)}_{n} & + & (1 - 2\gamma)^{2} T_{n,n-1} V^{(+)}_{n-1}  \nonumber \\
& = & 0 \\
(1+ 2\gamma)^{2} T_{n,n+1} V^{(-)}_{n+1} + [(1 - 4\gamma^{2})T_{n,n} - E ]
V^{(-)}_{n} & + &(1 - 2\gamma)^{2} T_{n,n-1} V^{(-)}_{n-1}  \nonumber \\
& = & \beta_0 <n|V|0>  \label{eq:recrel}
\end{eqnarray}

If in (\ref{eq:recrel}) the potential is switched off ($V_0 = 1$ and $z=1$
(or $\gamma = 0$)), the recurrence relations for $V_{n}^{(+)}$ and $%
V_{n}^{(-)}$ revert to these of $c_{n}^{(+)}$ and $c_{n}^{(-)}$, as is to be
expected.

%\input{biblio}

%\bibliography{alg_model}

\end{document}